\documentclass[twoside,12pt,draft,onecolumn]{IEEEtran} 
\usepackage{times,amsmath,amssymb,graphics,epsfig}

\begin{document}
\newcommand{\hp}{\mbox{\boldmath $D_{ec}^{\mbox{\unboldmath \scriptsize $\dg$}} D_{ec}$}}
\newcommand{\dm}{\mbox{\boldmath $D_{ec}$}} \newcommand{\dmm}{\mbox{\boldmath $D_{\mbox{\unboldmath \scriptsize $ec$}}$}}
\newcommand{\cw}{\mbox{\boldmath $c$}}
\newcommand{\cM}{\mbox{\boldmath $D_c$}}
\newcommand{\SM}{\mbox{\boldmath $S$}}
\newcommand{\sw}{\mbox{\boldmath $s$}}
\newcommand{\chiw}{\mbox{\boldmath $\chi$}}
\newcommand{\betaM}{\mbox{\boldmath $\beta$}}
\newcommand{\etaM}{\mbox{\boldmath $\eta$}}
\newcommand{\be}{\begin{equation}}
\newcommand{\ee}{\end{equation}}
\newcommand{\ew}{\mbox{\boldmath $e$}}
\newcommand{\eM}{\mbox{\boldmath $D_e$}}
\newcommand{\rf}[1]{\mbox{(\ref{#1})}}
\newcommand{\erfc}{\mbox{erfc}}
\newcommand{\diag}{\mbox{diag}}
\newcommand{\rank}{\mbox{rank}}
\newcommand{\tr}{\mbox{tr}}
\newcommand{\Prob}{\mbox{Pr}}
\newcommand{\bd}[1]{\mbox{\boldmath $#1$}}
\newcommand{\us}[1]{\mbox{\unboldmath \scriptsize $#1$}}
\newcommand{\ho}[1]{\mbox{\hollow #1}}
\newcommand{\df}{\stackrel{\mbox{\scriptsize def}}{=}}
\newcommand{\dg}{\dagger}
\newcommand{\laeq}{\stackrel{<}{_\sim }}
\newcommand{\bc}[2]{\mbox{$\left(\begin{array}{@{}c@{}}#1 \\ #2 \end{array}\right)$}}
\newtheorem{lemma}{\em Lemma}
\newtheorem{defn}{\em Definition}
\newtheorem{theorem}{\bf Theorem}
\newtheorem{corollary}{\em Corollary}
\newcommand{\asa}[2]{\mbox{$\left[\renewcommand{\arraystretch}{.5}\begin{array}{@{}c@{}c@{}} #1 & \,#2\\ #2^\ast & \,-#1^\ast  \end{array}\right]\renewcommand{\arraystretch}{1}$}}
\newcommand{\asb}[2]{\mbox{$\left[\renewcommand{\arraystretch}{.5}\begin{array}{@{}c@{}c@{}} #1 & \,#2\\ -#2^\ast & \,#1^\ast  \end{array}\right]\renewcommand{\arraystretch}{1}$}}
\newcommand{\mtrx}[4]{\mbox{$\left[\renewcommand{\arraystretch}{0.5}\begin{array}{@{}c@{}c@{}} #1 & \,#2\\ #3 & \,#4  \end{array}\right]\renewcommand{\arraystretch}{1}$}}

\newcommand{\tbb}[1]{\begin{table}[#1]
\caption{$2\!\times \!\!2$ matrices $\bd{C}_0\!, \ldots, \!\bd{C}_{31}$, form multidimensional space-time constellation, with subscript indices used in trellis branch labels. Entries of $\bd{C}_i$, $i\!=\!0, \ldots, \! 31$, represent  indices of  complex points from  a  4PSK constellation (\mbox{Table~\protect\ref{tab:qpsk}}). Each $\bd{C}_i$ defines   4PSK symbols  to be sent over  $L=2$ transmit antennas, during two consecutive complex symbol epochs.
}
\label{tab:superc}
\begin{center}
\begin{tabular}{|c|c|c|c|} \hline
$\bd{C}_0 $ \ldots $\bd{C}_7 $ & $\bd{C}_8 $ \ldots $\bd{C}_{15} $ & $\bd{C}_{16} $ \ldots $\bd{C}_{23} $ & $\bd{C}_{24} $ \ldots $\bd{C}_{31} $ \\ [7pt] \hline  \hline
$\left[\renewcommand{\arraystretch}{.7}\begin{array}{@{\:}c@{\:}c@{\:}} 0 & 3 \\ 0 & 1  \end{array}\right]$ &
$\left[\renewcommand{\arraystretch}{.7}\begin{array}{@{\:}c@{\:}c@{\:}} 2 & 3 \\ 0 & 3  \end{array}\right]$ &
$\left[\renewcommand{\arraystretch}{.7}\begin{array}{@{\:}c@{\:}c@{\:}} 0 & 1 \\ 0 & 3  \end{array}\right]$ &
$\left[\renewcommand{\arraystretch}{.7}\begin{array}{@{\:}c@{\:}c@{\:}} 2 & 1 \\ 0 & 1  \end{array}\right]$ \\ [7pt]
$\left[\renewcommand{\arraystretch}{.7}\begin{array}{@{\:}c@{\:}c@{\:}} 1 & 3 \\ 0 & 0  \end{array}\right]$ &
$\left[\renewcommand{\arraystretch}{.7}\begin{array}{@{\:}c@{\:}c@{\:}} 3 & 3 \\ 0 & 2  \end{array}\right]$ &
$\left[\renewcommand{\arraystretch}{.7}\begin{array}{@{\:}c@{\:}c@{\:}} 1 & 1 \\ 0 & 2  \end{array}\right]$ &
$\left[\renewcommand{\arraystretch}{.7}\begin{array}{@{\:}c@{\:}c@{\:}} 3 & 1 \\ 0 & 0  \end{array}\right]$ \\ [7pt]
$\left[\renewcommand{\arraystretch}{.7}\begin{array}{@{\:}c@{\:}c@{\:}} 0 & 1 \\ 2 & 1  \end{array}\right]$ &
$\left[\renewcommand{\arraystretch}{.7}\begin{array}{@{\:}c@{\:}c@{\:}} 2 & 1 \\ 2 & 3  \end{array}\right]$ &
$\left[\renewcommand{\arraystretch}{.7}\begin{array}{@{\:}c@{\:}c@{\:}} 0 & 3 \\ 2 & 3  \end{array}\right]$ &
$\left[\renewcommand{\arraystretch}{.7}\begin{array}{@{\:}c@{\:}c@{\:}} 2 & 3 \\ 2 & 1  \end{array}\right]$ \\ [7pt]
$\left[\renewcommand{\arraystretch}{.7}\begin{array}{@{\:}c@{\:}c@{\:}} 1 & 1 \\ 2 & 0  \end{array}\right]$ &
$\left[\renewcommand{\arraystretch}{.7}\begin{array}{@{\:}c@{\:}c@{\:}} 3 & 1 \\ 2 & 2  \end{array}\right]$ &
$\left[\renewcommand{\arraystretch}{.7}\begin{array}{@{\:}c@{\:}c@{\:}} 1 & 3 \\ 2 & 2  \end{array}\right]$ &
$\left[\renewcommand{\arraystretch}{.7}\begin{array}{@{\:}c@{\:}c@{\:}} 3 & 3 \\ 2 & 0  \end{array}\right]$ \\ [7pt]
$\left[\renewcommand{\arraystretch}{.7}\begin{array}{@{\:}c@{\:}c@{\:}} 0 & 2 \\ 1 & 1  \end{array}\right]$ &
$\left[\renewcommand{\arraystretch}{.7}\begin{array}{@{\:}c@{\:}c@{\:}} 2 & 2 \\ 1 & 3  \end{array}\right]$ &
$\left[\renewcommand{\arraystretch}{.7}\begin{array}{@{\:}c@{\:}c@{\:}} 0 & 0 \\ 1 & 3  \end{array}\right]$ &
$\left[\renewcommand{\arraystretch}{.7}\begin{array}{@{\:}c@{\:}c@{\:}} 2 & 0 \\ 1 & 1  \end{array}\right]$ \\ [7pt]
$\left[\renewcommand{\arraystretch}{.7}\begin{array}{@{\:}c@{\:}c@{\:}} 1 & 2 \\ 1 & 0  \end{array}\right]$ &
$\left[\renewcommand{\arraystretch}{.7}\begin{array}{@{\:}c@{\:}c@{\:}} 3 & 2 \\ 1 & 2  \end{array}\right]$ &
$\left[\renewcommand{\arraystretch}{.7}\begin{array}{@{\:}c@{\:}c@{\:}} 1 & 0 \\ 1 & 2  \end{array}\right]$ &
$\left[\renewcommand{\arraystretch}{.7}\begin{array}{@{\:}c@{\:}c@{\:}} 3 & 0 \\ 1 & 0  \end{array}\right]$ \\ [7pt]
$\left[\renewcommand{\arraystretch}{.7}\begin{array}{@{\:}c@{\:}c@{\:}} 0 & 0 \\ 3 & 1  \end{array}\right]$ &
$\left[\renewcommand{\arraystretch}{.7}\begin{array}{@{\:}c@{\:}c@{\:}} 2 & 0 \\ 3 & 3  \end{array}\right]$ &
$\left[\renewcommand{\arraystretch}{.7}\begin{array}{@{\:}c@{\:}c@{\:}} 0 & 2 \\ 3 & 3  \end{array}\right]$ &
$\left[\renewcommand{\arraystretch}{.7}\begin{array}{@{\:}c@{\:}c@{\:}} 2 & 2 \\ 3 & 1  \end{array}\right]$ \\ [7pt]
$\left[\renewcommand{\arraystretch}{.7}\begin{array}{@{\:}c@{\:}c@{\:}} 1 & 0 \\ 3 & 0  \end{array}\right]$ &
$\left[\renewcommand{\arraystretch}{.7}\begin{array}{@{\:}c@{\:}c@{\:}} 3 & 0 \\ 3 & 2  \end{array}\right]$ &
$\left[\renewcommand{\arraystretch}{.7}\begin{array}{@{\:}c@{\:}c@{\:}} 1 & 2 \\ 3 & 2  \end{array}\right]$ &
$\left[\renewcommand{\arraystretch}{.7}\begin{array}{@{\:}c@{\:}c@{\:}} 3 & 2 \\ 3 & 0  
\end{array}\renewcommand{\arraystretch}{1}\right]$  \\ [7pt] \hline
\end{tabular}
\end{center}
\end{table}
}


\newcommand{\tbbBIS}[1]{\begin{table}[#1]
\caption{The $2\!\times \! 2$ matrices $\bd{C}_i$, $i\!=\!0, \ldots , 31$, along with  relevant cosets ${\cal C}_l$ and corresponding uncoded bits, vs.\  number of states $q$.
}

\label{tab:superc}
\begin{center}
\begin{tabular}{@{\vline}c@{\vline}c@{\vline}c@{\vline}c@{\vline}c@{\vline}c@{\vline}c@{\vline}c@{\vline}c@{\vline}c@{\vline}c@{\vline}c@{\vline}}
\hline
\multicolumn{1}{@{\vline}c@{\vline}}{$i=0$} & \multicolumn{2}{c@{\vline}}{$q$} & \multicolumn{1}{c@{\vline}}{$i=8$} & \multicolumn{2}{c@{\vline}}{$q$} & \multicolumn{1}{c@{\vline}}{$i\!=\!16$} & \multicolumn{2}{c@{\vline}}{$q$} & \multicolumn{1}{c@{\vline}}{$i\!=\!24$} & \multicolumn{2}{c@{\vline}}{$q$} \\
\cline{2-3} \cline{5-6} \cline{8-9} \cline{11-12}
\multicolumn{1}{@{\vline}c@{\vline}}{$ \ldots 7$} & \multicolumn{1}{c@{\vline}}{8} & \multicolumn{1}{c@{\vline}}{16} & \multicolumn{1}{c@{\vline}}{$ \ldots 15$} & \multicolumn{1}{c@{\vline}}{8} & \multicolumn{1}{c@{\vline}}{16} & \multicolumn{1}{c@{\vline}}{$ \ldots 23$} & \multicolumn{1}{c@{\vline}}{8} & \multicolumn{1}{c@{\vline}}{16} & \multicolumn{1}{c@{\vline}}{$\ldots 31$} & \multicolumn{1}{c@{\vline}}{8} & \multicolumn{1}{c@{\vline}}{16}  \\
\hline
\hline
 & & & & & & & & & & & \\
$\left[\renewcommand{\arraystretch}{.7}\begin{array}{@{\:}c@{\:}c@{\:}} 1 & 3 \\ 0 & 0  \end{array}\right]$ &
$\renewcommand{\arraystretch}{.7}\begin{array}{@{\:}c@{\:}} {\cal C}_0 \\ 00  \end{array}$ &
$\renewcommand{\arraystretch}{.7}\begin{array}{@{\:}c@{\:}} {\cal C}_0 \\ 0  \end{array}$ &
$\left[\renewcommand{\arraystretch}{.7}\begin{array}{@{\:}c@{\:}c@{\:}} 3 & 3 \\ 0 & 2  \end{array}\right]$ &
$\renewcommand{\arraystretch}{.7}\begin{array}{@{\:}c@{\:}} {\cal C}_0 \\ 01  \end{array}$ &
$\renewcommand{\arraystretch}{.7}\begin{array}{@{\:}c@{\:}} {\cal C}_2 \\ 1  \end{array}$ &
$\left[\renewcommand{\arraystretch}{.7}\begin{array}{@{\:}c@{\:}c@{\:}} 3 & 1 \\ 0 & 0  \end{array}\right]$ &
$\renewcommand{\arraystretch}{.7}\begin{array}{@{\:}c@{\:}} {\cal C}_5 \\ 00  \end{array}$ &
$\renewcommand{\arraystretch}{.7}\begin{array}{@{\:}c@{\:}} {\cal C}_8 \\ 0  \end{array}$ &
$\left[\renewcommand{\arraystretch}{.7}\begin{array}{@{\:}c@{\:}c@{\:}} 1 & 1 \\ 0 & 2  \end{array}\right]$ &
$\renewcommand{\arraystretch}{.7}\begin{array}{@{\:}c@{\:}} {\cal C}_5 \\ 01  \end{array}$ &
$\renewcommand{\arraystretch}{.7}\begin{array}{@{\:}c@{\:}} {\cal C}_{10} \\ 1  \end{array}$   \\ [7pt]
$\left[\renewcommand{\arraystretch}{.7}\begin{array}{@{\:}c@{\:}c@{\:}} 1 & 2 \\ 1 & 0  \end{array}\right]$ &
$\renewcommand{\arraystretch}{.7}\begin{array}{@{\:}c@{\:}} {\cal C}_1 \\ 00  \end{array}$ &
$\renewcommand{\arraystretch}{.7}\begin{array}{@{\:}c@{\:}} {\cal C}_1 \\ 0  \end{array}$ &
$\left[\renewcommand{\arraystretch}{.7}\begin{array}{@{\:}c@{\:}c@{\:}} 3 & 2 \\ 1 & 2  \end{array}\right]$ &
$\renewcommand{\arraystretch}{.7}\begin{array}{@{\:}c@{\:}} {\cal C}_1 \\ 01  \end{array}$ &
$\renewcommand{\arraystretch}{.7}\begin{array}{@{\:}c@{\:}} {\cal C}_3 \\ 1  \end{array}$ &
$\left[\renewcommand{\arraystretch}{.7}\begin{array}{@{\:}c@{\:}c@{\:}} 3 & 0 \\ 1 & 0  \end{array}\right]$ &
$\renewcommand{\arraystretch}{.7}\begin{array}{@{\:}c@{\:}} {\cal C}_4 \\ 00  \end{array}$ &
$\renewcommand{\arraystretch}{.7}\begin{array}{@{\:}c@{\:}} {\cal C}_9 \\ 0  \end{array}$ &
$\left[\renewcommand{\arraystretch}{.7}\begin{array}{@{\:}c@{\:}c@{\:}} 1 & 0 \\ 1 & 2  \end{array}\right]$ &
$\renewcommand{\arraystretch}{.7}\begin{array}{@{\:}c@{\:}} {\cal C}_4 \\ 01  \end{array}$ &
$\renewcommand{\arraystretch}{.7}\begin{array}{@{\:}c@{\:}} {\cal C}_{11} \\ 1  \end{array}$   \\ [7pt]
$\left[\renewcommand{\arraystretch}{.7}\begin{array}{@{\:}c@{\:}c@{\:}} 1 & 1 \\ 2 & 0  \end{array}\right]$ &
$\renewcommand{\arraystretch}{.7}\begin{array}{@{\:}c@{\:}} {\cal C}_0 \\ 10  \end{array}$ &
$\renewcommand{\arraystretch}{.7}\begin{array}{@{\:}c@{\:}} {\cal C}_2 \\ 0  \end{array}$ &
$\left[\renewcommand{\arraystretch}{.7}\begin{array}{@{\:}c@{\:}c@{\:}} 3 & 1 \\ 2 & 2  \end{array}\right]$ &
$\renewcommand{\arraystretch}{.7}\begin{array}{@{\:}c@{\:}} {\cal C}_0 \\ 11  \end{array}$ &
$\renewcommand{\arraystretch}{.7}\begin{array}{@{\:}c@{\:}} {\cal C}_0 \\ 1  \end{array}$ &
$\left[\renewcommand{\arraystretch}{.7}\begin{array}{@{\:}c@{\:}c@{\:}} 3 & 3 \\ 2 & 0  \end{array}\right]$ &
$\renewcommand{\arraystretch}{.7}\begin{array}{@{\:}c@{\:}} {\cal C}_5 \\ 10  \end{array}$ &
$\renewcommand{\arraystretch}{.7}\begin{array}{@{\:}c@{\:}} {\cal C}_{10} \\ 0  \end{array}$ &
$\left[\renewcommand{\arraystretch}{.7}\begin{array}{@{\:}c@{\:}c@{\:}} 1 & 3 \\ 2 & 2  \end{array}\right]$ &
$\renewcommand{\arraystretch}{.7}\begin{array}{@{\:}c@{\:}} {\cal C}_5 \\ 11  \end{array}$ &
$\renewcommand{\arraystretch}{.7}\begin{array}{@{\:}c@{\:}} {\cal C}_8 \\ 1  \end{array}$   \\ [7pt]
$\left[\renewcommand{\arraystretch}{.7}\begin{array}{@{\:}c@{\:}c@{\:}} 1 & 0 \\ 3 & 0  \end{array}\right]$ &
$\renewcommand{\arraystretch}{.7}\begin{array}{@{\:}c@{\:}} {\cal C}_1 \\ 10  \end{array}$ &
$\renewcommand{\arraystretch}{.7}\begin{array}{@{\:}c@{\:}} {\cal C}_3 \\ 0  \end{array}$ &
$\left[\renewcommand{\arraystretch}{.7}\begin{array}{@{\:}c@{\:}c@{\:}} 3 & 0 \\ 3 & 2  \end{array}\right]$ &
$\renewcommand{\arraystretch}{.7}\begin{array}{@{\:}c@{\:}} {\cal C}_1 \\ 11  \end{array}$ &
$\renewcommand{\arraystretch}{.7}\begin{array}{@{\:}c@{\:}} {\cal C}_1 \\ 1  \end{array}$ &
$\left[\renewcommand{\arraystretch}{.7}\begin{array}{@{\:}c@{\:}c@{\:}} 3 & 2 \\ 3 & 0  \end{array}\right]$ &
$\renewcommand{\arraystretch}{.7}\begin{array}{@{\:}c@{\:}} {\cal C}_4 \\ 10  \end{array}$ &
$\renewcommand{\arraystretch}{.7}\begin{array}{@{\:}c@{\:}} {\cal C}_{11} \\ 0  \end{array}$ &
$\left[\renewcommand{\arraystretch}{.7}\begin{array}{@{\:}c@{\:}c@{\:}} 1 & 2 \\ 3 & 2  \end{array}\right]$ &
$\renewcommand{\arraystretch}{.7}\begin{array}{@{\:}c@{\:}} {\cal C}_4 \\ 11  \end{array}$ &
$\renewcommand{\arraystretch}{.7}\begin{array}{@{\:}c@{\:}} {\cal C}_9 \\ 1  \end{array}$   \\ [7pt]
$\left[\renewcommand{\arraystretch}{.7}\begin{array}{@{\:}c@{\:}c@{\:}} 0 & 3 \\ 0 & 1  \end{array}\right]$ &
$\renewcommand{\arraystretch}{.7}\begin{array}{@{\:}c@{\:}} {\cal C}_3 \\ 00  \end{array}$ &
$\renewcommand{\arraystretch}{.7}\begin{array}{@{\:}c@{\:}} {\cal C}_5 \\ 0  \end{array}$ &
$\left[\renewcommand{\arraystretch}{.7}\begin{array}{@{\:}c@{\:}c@{\:}} 2 & 3 \\ 0 & 3  \end{array}\right]$ &
$\renewcommand{\arraystretch}{.7}\begin{array}{@{\:}c@{\:}} {\cal C}_3 \\ 01  \end{array}$ &
$\renewcommand{\arraystretch}{.7}\begin{array}{@{\:}c@{\:}} {\cal C}_7 \\ 1  \end{array}$ &
$\left[\renewcommand{\arraystretch}{.7}\begin{array}{@{\:}c@{\:}c@{\:}} 2 & 1 \\ 0 & 1  \end{array}\right]$ &
$\renewcommand{\arraystretch}{.7}\begin{array}{@{\:}c@{\:}} {\cal C}_6 \\ 00  \end{array}$ &
$\renewcommand{\arraystretch}{.7}\begin{array}{@{\:}c@{\:}} {\cal C}_{13} \\ 0  \end{array}$ &
$\left[\renewcommand{\arraystretch}{.7}\begin{array}{@{\:}c@{\:}c@{\:}} 0 & 1 \\ 0 & 3  \end{array}\right]$ &
$\renewcommand{\arraystretch}{.7}\begin{array}{@{\:}c@{\:}} {\cal C}_6 \\ 01  \end{array}$ &
$\renewcommand{\arraystretch}{.7}\begin{array}{@{\:}c@{\:}} {\cal C}_{15} \\ 1  \end{array}$   \\ [7pt]
$\left[\renewcommand{\arraystretch}{.7}\begin{array}{@{\:}c@{\:}c@{\:}} 0 & 2 \\ 1 & 1  \end{array}\right]$ &
$\renewcommand{\arraystretch}{.7}\begin{array}{@{\:}c@{\:}} {\cal C}_2 \\ 00  \end{array}$ &
$\renewcommand{\arraystretch}{.7}\begin{array}{@{\:}c@{\:}} {\cal C}_4 \\ 0  \end{array}$ &
$\left[\renewcommand{\arraystretch}{.7}\begin{array}{@{\:}c@{\:}c@{\:}} 2 & 2 \\ 1 & 3  \end{array}\right]$ &
$\renewcommand{\arraystretch}{.7}\begin{array}{@{\:}c@{\:}} {\cal C}_2 \\ 01  \end{array}$ &
$\renewcommand{\arraystretch}{.7}\begin{array}{@{\:}c@{\:}} {\cal C}_6 \\ 1  \end{array}$ &
$\left[\renewcommand{\arraystretch}{.7}\begin{array}{@{\:}c@{\:}c@{\:}} 2 & 0 \\ 1 & 1  \end{array}\right]$ &
$\renewcommand{\arraystretch}{.7}\begin{array}{@{\:}c@{\:}} {\cal C}_7 \\ 00  \end{array}$ &
$\renewcommand{\arraystretch}{.7}\begin{array}{@{\:}c@{\:}} {\cal C}_{12} \\ 0  \end{array}$ &
$\left[\renewcommand{\arraystretch}{.7}\begin{array}{@{\:}c@{\:}c@{\:}} 0 & 0 \\ 1 & 3  \end{array}\right]$ &
$\renewcommand{\arraystretch}{.7}\begin{array}{@{\:}c@{\:}} {\cal C}_7 \\ 01  \end{array}$ &
$\renewcommand{\arraystretch}{.7}\begin{array}{@{\:}c@{\:}} {\cal C}_{14} \\ 1  \end{array}$   \\ [7pt]
$\left[\renewcommand{\arraystretch}{.7}\begin{array}{@{\:}c@{\:}c@{\:}} 0 & 1 \\ 2 & 1  \end{array}\right]$ &
$\renewcommand{\arraystretch}{.7}\begin{array}{@{\:}c@{\:}} {\cal C}_3 \\ 10  \end{array}$ &
$\renewcommand{\arraystretch}{.7}\begin{array}{@{\:}c@{\:}} {\cal C}_7 \\ 0  \end{array}$ &
$\left[\renewcommand{\arraystretch}{.7}\begin{array}{@{\:}c@{\:}c@{\:}} 2 & 1 \\ 2 & 3  \end{array}\right]$ &
$\renewcommand{\arraystretch}{.7}\begin{array}{@{\:}c@{\:}} {\cal C}_3 \\ 11  \end{array}$ &
$\renewcommand{\arraystretch}{.7}\begin{array}{@{\:}c@{\:}} {\cal C}_5 \\ 1  \end{array}$ &
$\left[\renewcommand{\arraystretch}{.7}\begin{array}{@{\:}c@{\:}c@{\:}} 2 & 3 \\ 2 & 1  \end{array}\right]$ &
$\renewcommand{\arraystretch}{.7}\begin{array}{@{\:}c@{\:}} {\cal C}_6 \\ 10  \end{array}$ &
$\renewcommand{\arraystretch}{.7}\begin{array}{@{\:}c@{\:}} {\cal C}_{15} \\ 0  \end{array}$ &
$\left[\renewcommand{\arraystretch}{.7}\begin{array}{@{\:}c@{\:}c@{\:}} 0 & 3 \\ 2 & 3  \end{array}\right]$ &
$\renewcommand{\arraystretch}{.7}\begin{array}{@{\:}c@{\:}} {\cal C}_6 \\ 11  \end{array}$ &
$\renewcommand{\arraystretch}{.7}\begin{array}{@{\:}c@{\:}} {\cal C}_{13} \\ 1  \end{array}$  \\ [7pt]
$\left[\renewcommand{\arraystretch}{.7}\begin{array}{@{\:}c@{\:}c@{\:}} 0 & 0 \\ 3 & 1  \end{array}\right]$ &
$\renewcommand{\arraystretch}{.7}\begin{array}{@{\:}c@{\:}} {\cal C}_2 \\ 10  \end{array}$ &
$\renewcommand{\arraystretch}{.7}\begin{array}{@{\:}c@{\:}} {\cal C}_6 \\ 0  \end{array}$ &
$\left[\renewcommand{\arraystretch}{.7}\begin{array}{@{\:}c@{\:}c@{\:}} 2 & 0 \\ 3 & 3  \end{array}\right]$ &
$\renewcommand{\arraystretch}{.7}\begin{array}{@{\:}c@{\:}} {\cal C}_2 \\ 11  \end{array}$ &
$\renewcommand{\arraystretch}{.7}\begin{array}{@{\:}c@{\:}} {\cal C}_4 \\ 1  \end{array}$ &
$\left[\renewcommand{\arraystretch}{.7}\begin{array}{@{\:}c@{\:}c@{\:}} 2 & 2 \\ 3 & 1  \end{array}\right]$ &
$\renewcommand{\arraystretch}{.7}\begin{array}{@{\:}c@{\:}} {\cal C}_7 \\ 10  \end{array}$ &
$\renewcommand{\arraystretch}{.7}\begin{array}{@{\:}c@{\:}} {\cal C}_{14} \\ 0  \end{array}$ &
$\left[\renewcommand{\arraystretch}{.7}\begin{array}{@{\:}c@{\:}c@{\:}} 0 & 2 \\ 3 & 3  \end{array}\right]$ &
$\renewcommand{\arraystretch}{.7}\begin{array}{@{\:}c@{\:}} {\cal C}_7 \\ 11  \end{array}$ &
$\renewcommand{\arraystretch}{.7}\begin{array}{@{\:}c@{\:}} {\cal C}_{12} \\ 1  \end{array}\renewcommand{\arraystretch}{1}$  \\ [7pt] \hline
\end{tabular}
\end{center}
\end{table}
}



\newcommand{\multiDIMconstellREORDandREORDbo}[1]{\begin{table}[#1]
\caption{Two different orderings of the $2\!\times \!\!2$ matrices $\bd{C}_0\!, \ldots, \!\bd{C}_{31}$, which form a multidimensional space-time constellation; subscript indices $i$ are used in trellis branch labels. Entries of $\bd{C}_i$, $i\!=\!0, \ldots, \! 31$, represent  indices of  complex points from the 4PSK constellation. Each $\bd{C}_i$ defines 4PSK symbols  to be sent over  $L=2$ transmit antennas, during two consecutive complex symbol epochs.
}
\label{tab:supercREORDandREORDbo}
\begin{center}
\begin{tabular}{@{\vline}c@{\vline}c@{\vline}c@{\vline}c@{\vline}c@{\vline}c@{\vline}c@{\vline}c@{\vline}c@{\vline}c@{\vline}c@{\vline}c@{\vline}}
\hline
\multicolumn{1}{@{\vline}c@{\vline}}{$\bd{C}_i$} & \multicolumn{2}{c@{\vline}}{$i$} & \multicolumn{1}{c@{\vline}}{$\bd{C}_i$} & \multicolumn{2}{c@{\vline}}{$i$} & \multicolumn{1}{c@{\vline}}{$\bd{C}_i$} & \multicolumn{2}{c@{\vline}}{$i$} & \multicolumn{1}{c@{\vline}}{$\bd{C}_i$} & \multicolumn{2}{c@{\vline}}{$i$} \\
\cline{2-3} \cline{5-6} \cline{8-9} \cline{11-12}
\multicolumn{1}{@{\vline}c@{\vline}}{} & \multicolumn{1}{c@{\vline}}{\scriptsize new} & \multicolumn{1}{c@{\vline}}{\scriptsize old} & \multicolumn{1}{c@{\vline}}{} & \multicolumn{1}{c@{\vline}}{\scriptsize new} & \multicolumn{1}{c@{\vline}}{\scriptsize old} & \multicolumn{1}{c@{\vline}}{} & \multicolumn{1}{c@{\vline}}{\scriptsize new} & \multicolumn{1}{c@{\vline}}{\scriptsize old} & \multicolumn{1}{c@{\vline}}{} & \multicolumn{1}{c@{\vline}}{\scriptsize new} & \multicolumn{1}{c@{\vline}}{\scriptsize old}  \\
\hline
\hline
 & & & & & & & & & & & \\
$\left[\renewcommand{\arraystretch}{.7}\begin{array}{@{\:}c@{\:}c@{\:}} 0 & 3 \\ 0 & 1  \end{array}\right]$ &
0  & 
0  & 
$\left[\renewcommand{\arraystretch}{.7}\begin{array}{@{\:}c@{\:}c@{\:}} 2 & 3 \\ 0 & 3  \end{array}\right]$ &
12 & 
8  & 
$\left[\renewcommand{\arraystretch}{.7}\begin{array}{@{\:}c@{\:}c@{\:}} 0 & 1 \\ 0 & 3  \end{array}\right]$ &
16 & 
16 & 
$\left[\renewcommand{\arraystretch}{.7}\begin{array}{@{\:}c@{\:}c@{\:}} 2 & 1 \\ 0 & 1  \end{array}\right]$ &
28 & 
24 \\ [7pt] 
$\left[\renewcommand{\arraystretch}{.7}\begin{array}{@{\:}c@{\:}c@{\:}} 0 & 2 \\ 1 & 1  \end{array}\right]$ &
1  & 
1  & 
$\left[\renewcommand{\arraystretch}{.7}\begin{array}{@{\:}c@{\:}c@{\:}} 2 & 2 \\ 1 & 3  \end{array}\right]$ &
13 & 
9  & 
$\left[\renewcommand{\arraystretch}{.7}\begin{array}{@{\:}c@{\:}c@{\:}} 0 & 0 \\ 1 & 3  \end{array}\right]$ &
17 & 
17 & 
$\left[\renewcommand{\arraystretch}{.7}\begin{array}{@{\:}c@{\:}c@{\:}} 2 & 0 \\ 1 & 1  \end{array}\right]$ &
29 & 
25 \\ [7pt] 
$\left[\renewcommand{\arraystretch}{.7}\begin{array}{@{\:}c@{\:}c@{\:}} 0 & 1 \\ 2 & 1  \end{array}\right]$ &
3  & 
2  & 
$\left[\renewcommand{\arraystretch}{.7}\begin{array}{@{\:}c@{\:}c@{\:}} 2 & 1 \\ 2 & 3  \end{array}\right]$ &
15 & 
10 & 
$\left[\renewcommand{\arraystretch}{.7}\begin{array}{@{\:}c@{\:}c@{\:}} 0 & 3 \\ 2 & 3  \end{array}\right]$ &
19 & 
18 & 
$\left[\renewcommand{\arraystretch}{.7}\begin{array}{@{\:}c@{\:}c@{\:}} 2 & 3 \\ 2 & 1  \end{array}\right]$ &
31 & 
26 \\ [7pt] 
$\left[\renewcommand{\arraystretch}{.7}\begin{array}{@{\:}c@{\:}c@{\:}} 0 & 0 \\ 3 & 1  \end{array}\right]$ &
2  & 
3  & 
$\left[\renewcommand{\arraystretch}{.7}\begin{array}{@{\:}c@{\:}c@{\:}} 2 & 0 \\ 3 & 3  \end{array}\right]$ &
14 & 
11 & 
$\left[\renewcommand{\arraystretch}{.7}\begin{array}{@{\:}c@{\:}c@{\:}} 0 & 2 \\ 3 & 3  \end{array}\right]$ &
18 & 
19 & 
$\left[\renewcommand{\arraystretch}{.7}\begin{array}{@{\:}c@{\:}c@{\:}} 2 & 2 \\ 3 & 1  \end{array}\right]$ &
30 & 
27 \\ [7pt] 
$\left[\renewcommand{\arraystretch}{.7}\begin{array}{@{\:}c@{\:}c@{\:}} 1 & 3 \\ 0 & 0  \end{array}\right]$ &
4  & 
4  & 
$\left[\renewcommand{\arraystretch}{.7}\begin{array}{@{\:}c@{\:}c@{\:}} 3 & 3 \\ 0 & 2  \end{array}\right]$ &
8  & 
12 & 
$\left[\renewcommand{\arraystretch}{.7}\begin{array}{@{\:}c@{\:}c@{\:}} 1 & 1 \\ 0 & 2  \end{array}\right]$ &
20 & 
20 & 
$\left[\renewcommand{\arraystretch}{.7}\begin{array}{@{\:}c@{\:}c@{\:}} 3 & 1 \\ 0 & 0  \end{array}\right]$ &
24 & 
28 \\ [7pt] 
$\left[\renewcommand{\arraystretch}{.7}\begin{array}{@{\:}c@{\:}c@{\:}} 1 & 2 \\ 1 & 0  \end{array}\right]$ &
5  & 
5  & 
$\left[\renewcommand{\arraystretch}{.7}\begin{array}{@{\:}c@{\:}c@{\:}} 3 & 2 \\ 1 & 2  \end{array}\right]$ &
9  & 
13 & 
$\left[\renewcommand{\arraystretch}{.7}\begin{array}{@{\:}c@{\:}c@{\:}} 1 & 0 \\ 1 & 2  \end{array}\right]$ &
21 & 
21 & 
$\left[\renewcommand{\arraystretch}{.7}\begin{array}{@{\:}c@{\:}c@{\:}} 3 & 0 \\ 1 & 0  \end{array}\right]$ &
25 & 
29 \\ [7pt] 
$\left[\renewcommand{\arraystretch}{.7}\begin{array}{@{\:}c@{\:}c@{\:}} 1 & 1 \\ 2 & 0  \end{array}\right]$ &
7  & 
6  & 
$\left[\renewcommand{\arraystretch}{.7}\begin{array}{@{\:}c@{\:}c@{\:}} 3 & 1 \\ 2 & 2  \end{array}\right]$ &
11 & 
14 & 
$\left[\renewcommand{\arraystretch}{.7}\begin{array}{@{\:}c@{\:}c@{\:}} 1 & 3 \\ 2 & 2  \end{array}\right]$ &
23 & 
22 & 
$\left[\renewcommand{\arraystretch}{.7}\begin{array}{@{\:}c@{\:}c@{\:}} 3 & 3 \\ 2 & 0  \end{array}\right]$ &
27 & 
30 \\ [7pt] 
$\left[\renewcommand{\arraystretch}{.7}\begin{array}{@{\:}c@{\:}c@{\:}} 1 & 0 \\ 3 & 0  \end{array}\right]$ &
6  & 
7  & 
$\left[\renewcommand{\arraystretch}{.7}\begin{array}{@{\:}c@{\:}c@{\:}} 3 & 0 \\ 3 & 2  \end{array}\right]$ &
10 & 
15 & 
$\left[\renewcommand{\arraystretch}{.7}\begin{array}{@{\:}c@{\:}c@{\:}} 1 & 2 \\ 3 & 2  \end{array}\right]$ &
22 & 
23 & 
$\left[\renewcommand{\arraystretch}{.7}\begin{array}{@{\:}c@{\:}c@{\:}} 3 & 2 \\ 3 & 0  \end{array}\right]$ &
26 & 
31 \\ [7pt] \hline 
\end{tabular}
\end{center}
\end{table}
}


\newcommand{\multiDIMconstellREORDbo}[1]{\begin{table}[#1]
\caption{$2\!\times \!\!2$ matrices $\bd{C}_0\!, \ldots, \!\bd{C}_{31}$, form multidimensional space-time constellation, with subscript indices used in trellis branch labels. Entries of $\bd{C}_i$, $i\!=\!0, \ldots, \! 31$, represent  indices of  complex points from the 4PSK constellation. Each $\bd{C}_i$ defines 4PSK symbols  to be sent over  $L=2$ transmit antennas, during two consecutive complex symbol epochs.
}
\label{tab:supercREORDbo} 
\begin{center}
\begin{tabular}{|c|c|c|c|} \hline
$\bd{C}_0 $ \ldots $\bd{C}_7 $ & $\bd{C}_8 $ \ldots $\bd{C}_{15} $ & $\bd{C}_{16} $ \ldots $\bd{C}_{23} $ & $\bd{C}_{24} $ \ldots $\bd{C}_{31} $ \\ [1pt] \hline  \hline
$\left[\renewcommand{\arraystretch}{.7}\begin{array}{@{\:}c@{\:}c@{\:}} 0 & 3 \\ 0 & 1  \end{array}\right]$ &
$\left[\renewcommand{\arraystretch}{.7}\begin{array}{@{\:}c@{\:}c@{\:}} 3 & 3 \\ 0 & 2  \end{array}\right]$ &
$\left[\renewcommand{\arraystretch}{.7}\begin{array}{@{\:}c@{\:}c@{\:}} 0 & 1 \\ 0 & 3  \end{array}\right]$ &
$\left[\renewcommand{\arraystretch}{.7}\begin{array}{@{\:}c@{\:}c@{\:}} 3 & 1 \\ 0 & 0  \end{array}\right]$ \\ [5pt]
$\left[\renewcommand{\arraystretch}{.7}\begin{array}{@{\:}c@{\:}c@{\:}} 0 & 2 \\ 1 & 1  \end{array}\right]$ &
$\left[\renewcommand{\arraystretch}{.7}\begin{array}{@{\:}c@{\:}c@{\:}} 3 & 2 \\ 1 & 2  \end{array}\right]$ &
$\left[\renewcommand{\arraystretch}{.7}\begin{array}{@{\:}c@{\:}c@{\:}} 0 & 0 \\ 1 & 3  \end{array}\right]$ &
$\left[\renewcommand{\arraystretch}{.7}\begin{array}{@{\:}c@{\:}c@{\:}} 3 & 0 \\ 1 & 0  \end{array}\right]$ \\ [5pt]
$\left[\renewcommand{\arraystretch}{.7}\begin{array}{@{\:}c@{\:}c@{\:}} 0 & 0 \\ 3 & 1  \end{array}\right]$ &
$\left[\renewcommand{\arraystretch}{.7}\begin{array}{@{\:}c@{\:}c@{\:}} 3 & 0 \\ 3 & 2  \end{array}\right]$ &
$\left[\renewcommand{\arraystretch}{.7}\begin{array}{@{\:}c@{\:}c@{\:}} 0 & 2 \\ 3 & 3  \end{array}\right]$ &
$\left[\renewcommand{\arraystretch}{.7}\begin{array}{@{\:}c@{\:}c@{\:}} 3 & 2 \\ 3 & 0  \end{array}\right]$ \\ [5pt]
$\left[\renewcommand{\arraystretch}{.7}\begin{array}{@{\:}c@{\:}c@{\:}} 0 & 1 \\ 2 & 1  \end{array}\right]$ &
$\left[\renewcommand{\arraystretch}{.7}\begin{array}{@{\:}c@{\:}c@{\:}} 3 & 1 \\ 2 & 2  \end{array}\right]$ &
$\left[\renewcommand{\arraystretch}{.7}\begin{array}{@{\:}c@{\:}c@{\:}} 0 & 3 \\ 2 & 3  \end{array}\right]$ &
$\left[\renewcommand{\arraystretch}{.7}\begin{array}{@{\:}c@{\:}c@{\:}} 3 & 3 \\ 2 & 0  \end{array}\right]$ \\ [5pt]
$\left[\renewcommand{\arraystretch}{.7}\begin{array}{@{\:}c@{\:}c@{\:}} 1 & 3 \\ 0 & 0  \end{array}\right]$ &
$\left[\renewcommand{\arraystretch}{.7}\begin{array}{@{\:}c@{\:}c@{\:}} 2 & 3 \\ 0 & 3  \end{array}\right]$ &
$\left[\renewcommand{\arraystretch}{.7}\begin{array}{@{\:}c@{\:}c@{\:}} 1 & 1 \\ 0 & 2  \end{array}\right]$ &
$\left[\renewcommand{\arraystretch}{.7}\begin{array}{@{\:}c@{\:}c@{\:}} 2 & 1 \\ 0 & 1  \end{array}\right]$ \\ [5pt]
$\left[\renewcommand{\arraystretch}{.7}\begin{array}{@{\:}c@{\:}c@{\:}} 1 & 2 \\ 1 & 0  \end{array}\right]$ &
$\left[\renewcommand{\arraystretch}{.7}\begin{array}{@{\:}c@{\:}c@{\:}} 2 & 2 \\ 1 & 3  \end{array}\right]$ &
$\left[\renewcommand{\arraystretch}{.7}\begin{array}{@{\:}c@{\:}c@{\:}} 1 & 0 \\ 1 & 2  \end{array}\right]$ &
$\left[\renewcommand{\arraystretch}{.7}\begin{array}{@{\:}c@{\:}c@{\:}} 2 & 0 \\ 1 & 1  \end{array}\right]$ \\ [5pt]
$\left[\renewcommand{\arraystretch}{.7}\begin{array}{@{\:}c@{\:}c@{\:}} 1 & 0 \\ 3 & 0  \end{array}\right]$ &
$\left[\renewcommand{\arraystretch}{.7}\begin{array}{@{\:}c@{\:}c@{\:}} 2 & 0 \\ 3 & 3  \end{array}\right]$ &
$\left[\renewcommand{\arraystretch}{.7}\begin{array}{@{\:}c@{\:}c@{\:}} 1 & 2 \\ 3 & 2  \end{array}\right]$ &
$\left[\renewcommand{\arraystretch}{.7}\begin{array}{@{\:}c@{\:}c@{\:}} 2 & 2 \\ 3 & 1  \end{array}\right]$ \\ [5pt]
$\left[\renewcommand{\arraystretch}{.7}\begin{array}{@{\:}c@{\:}c@{\:}} 1 & 1 \\ 2 & 0  \end{array}\right]$ &
$\left[\renewcommand{\arraystretch}{.7}\begin{array}{@{\:}c@{\:}c@{\:}} 2 & 1 \\ 2 & 3  \end{array}\right]$ &
$\left[\renewcommand{\arraystretch}{.7}\begin{array}{@{\:}c@{\:}c@{\:}} 1 & 3 \\ 2 & 2  \end{array}\right]$ &
$\left[\renewcommand{\arraystretch}{.7}\begin{array}{@{\:}c@{\:}c@{\:}} 2 & 3 \\ 2 & 1  
\end{array}\renewcommand{\arraystretch}{1}\right]$  \\ [4pt] \hline
\end{tabular}
\end{center}
\end{table}
}



\newcommand{\tbQPSK}[1]{\begin{table}[#1]
\caption{Indexing for the 4PSK constellation points.}
\label{tab:qpsk}
\begin{center}
\begin{tabular}{|c|c|c|c|} \hline
$s_0 $  & $s_1 $  & $s_2 $  & $s_3 $  \\ [7pt] \hline  \hline
$\frac{1}{\sqrt{2}}+j \frac{1}{\sqrt{2}}$ &
$-\frac{1}{\sqrt{2}}+j \frac{1}{\sqrt{2}}$ &
$-\frac{1}{\sqrt{2}}-j \frac{1}{\sqrt{2}}$ &
$\frac{1}{\sqrt{2}}-j \frac{1}{\sqrt{2}}$ \\ [7pt]
 \hline
\end{tabular}
\end{center}
\end{table}
}
\newcommand{\tbbQPSK}[1]{\begin{table}[#1]
\caption{Indexing for the 4PSK constellation points.}
\label{tab:qpsk}
\begin{center}
\begin{tabular}{|c|c|c|c|} \hline
$s_0 $  & $s_1 $  & $s_2 $  & $s_3 $  \\ [1pt] \hline  \hline
$2^{-\frac{1}{2}}(1+j) $ &
$2^{-\frac{1}{2}}(-1+j) $ &
$2^{-\frac{1}{2}}(-1-j)$ &
$2^{-\frac{1}{2}}(1-j)$ \\ [1pt]
 \hline
\end{tabular}
\end{center}
\end{table}
}
\newcommand{\trellisfig}[2]{
\setlength{\unitlength}{#1in}
\begin{picture}(34,16)
\put(29,15){\line(3,0){6}}
\put(29,15){\line(3,-1){6}}
\put(29,15){\line(3,-2){6}}
\put(29,15){\line(1,-1){6}}
\put(29,13){\line(1,-1){6}}
\put(29,13){\line(3,-4){6}}
\put(29,13){\line(3,-5){6}}
\put(29,13){\line(1,-2){6}}

\put(29,11){\line(3,2){6}}
\put(29,11){\line(3,1){6}}
\put(29,11){\line(3,0){6}}
\put(29,11){\line(3,-1){6}}
\put(29,9){\line(3,-1){6}}
\put(29,9){\line(3,-2){6}}
\put(29,9){\line(1,-1){6}}
\put(29,9){\line(3,-4){6}}

\put(29,7){\line(3,4){6}}
\put(29,7){\line(1,1){6}}
\put(29,7){\line(3,2){6}}
\put(29,7){\line(3,1){6}}
\put(29,5){\line(3,1){6}}
\put(29,5){\line(3,0){6}}
\put(29,5){\line(3,-1){6}}
\put(29,5){\line(3,-2){6}}

\put(29,3){\line(1,2){6}}
\put(29,3){\line(3,5){6}}
\put(29,3){\line(3,4){6}}
\put(29,3){\line(1,1){6}}
\put(29,1){\line(1,1){6}}
\put(29,1){\line(3,2){6}}
\put(29,1){\line(3,1){6}}
\put(29,1){\line(3,0){6}}
\font\ff=#2
\put(0,14.5){\ff (0,10,2,8),(1,11,3,9),(5,15,7,13),(4,14,6,12)}
\put(0,12.5){\ff (17,27,19,25),(16,26,18,24),(20,30,22,28),(21,31,23,29)}
\put(0,10.5){\ff (5,15,7,13),(4,14,6,12), (0,10,2,8), (1,11,3,9)}
\put(0,8.5){\ff (20,30,22,28),(21,31,23,29), (17,27,19,25), (16,26,18,24)}

\put(0,6.5){\ff (1,11,3,9),(5,15,7,13),(4,14,6,12),(0,10,2,8)}
\put(0,4.5){\ff (16,26,18,24),(20,30,22,28),(21,31,23,29),(17,27,19,25)}
\put(0,2.5){\ff (4,14,6,12),(0,10,2,8),(1,11,3,9),(5,15,7,13)}
\put(0,0.5){\ff (21,31,23,29),(17,27,19,25),(16,26,18,24),(20,30,22,28)}
\put(29,0){\mbox{\ff \it n}}
\put(34,0){\mbox{\ff {\it n}+2}}
\put(28.5,1.2){\ff 7}
\put(28.5,3.2){\ff 6}
\put(28.5,5.2){\ff 5}
\put(28.5,7.2){\ff 4}
\put(28.5,9.2){\ff 3}
\put(28.5,11.2){\ff 2}
\put(28.5,13.2){\ff 1}
\put(28.5,15.2){\ff 0}
\end{picture}
}

\newcommand{\trellisfigfirstRSeightST}[2]{
\setlength{\unitlength}{#1in}
\begin{picture}(34,16)
\put(29,15){\line(3,0){6}}
\put(29,15){\line(3,-1){6}}
\put(29,15){\line(3,-2){6}}
\put(29,15){\line(1,-1){6}}
\put(29,13){\line(1,-1){6}}
\put(29,13){\line(3,-4){6}}
\put(29,13){\line(3,-5){6}}
\put(29,13){\line(1,-2){6}}

\put(29,11){\line(3,2){6}}
\put(29,11){\line(3,1){6}}
\put(29,11){\line(3,0){6}}
\put(29,11){\line(3,-1){6}}
\put(29,9){\line(3,-1){6}}
\put(29,9){\line(3,-2){6}}
\put(29,9){\line(1,-1){6}}
\put(29,9){\line(3,-4){6}}

\put(29,7){\line(3,4){6}}
\put(29,7){\line(1,1){6}}
\put(29,7){\line(3,2){6}}
\put(29,7){\line(3,1){6}}
\put(29,5){\line(3,1){6}}
\put(29,5){\line(3,0){6}}
\put(29,5){\line(3,-1){6}}
\put(29,5){\line(3,-2){6}}

\put(29,3){\line(1,2){6}}
\put(29,3){\line(3,5){6}}
\put(29,3){\line(3,4){6}}
\put(29,3){\line(1,1){6}}
\put(29,1){\line(1,1){6}}
\put(29,1){\line(3,2){6}}
\put(29,1){\line(3,1){6}}
\put(29,1){\line(3,0){6}}
\font\ff=CMR7 
\put(0,14.5){\ff (0,8,2,10),(4,12,6,14),(5,13,7,15),(1,9,3,11)}
\put(0,12.5){\ff (20,28,22,30),(16,24,18,26),(17,25,19,27),(21,29,23,31)}
\put(0,10.5){\ff (4,12,6,14),(0,8,2,10),(1,9,3,11),(5,13,7,15)}
\put(0,8.5){\ff (16,24,18,26),(20,28,22,30),(21,29,23,31),(17,25,19,27)}

\put(0,6.5){\ff (5,13,7,15),(1,9,3,11),(0,8,2,10),(4,12,6,14)}
\put(0,4.5){\ff (17,25,19,27),(21,29,23,31),(20,28,22,30),(16,24,18,26)}
\put(0,2.5){\ff (1,9,3,11),(5,13,7,15),(4,12,6,14),(0,8,2,10)}
\put(0,0.5){\ff (21,29,23,31),(17,25,19,27),(16,24,18,26),(20,28,22,30)}
\put(29,0){\mbox{\ff \it n}}
\put(33,0){\mbox{\ff {\it n}+2}}
\font\ff=#2
\put(26.8,0.5){\ff {{7}}}
\put(26.8,2.5){\ff {{6}}}
\put(26.8,4.5){\ff {{5}}}
\put(26.8,6.5){\ff {{4}}}
\put(26.8,8.5){\ff {{3}}}
\put(26.8,10.5){\ff {{2}}}
\put(26.8,12.50){\ff {{1}}}
\put(26.8,14.5){\ff {{0}}}
\put(27.5,0){\line(0,1){16}}
\font\ff=CMR7 
\put(27.7,1.2){\ff 3}
\put(28.3,1.0){\ff 1}
\put(28.3,0.3){\ff 0}
\put(27.7,0.0){\ff 2}

\put(27.7,3.2){\ff 1}
\put(28.3,3.0){\ff 3}
\put(28.3,2.2){\ff 2}
\put(27.7,2.0){\ff 0}

\put(27.7,5.2){\ff 1}
\put(28.3,5.0){\ff 3}
\put(28.3,4.2){\ff 2}
\put(27.7,4.0){\ff 0}

\put(27.7,7.2){\ff 3}
\put(28.3,7.0){\ff 1}
\put(28.3,6.2){\ff 0}
\put(27.7,6.0){\ff 2}

\put(27.7,9.2){\ff 0}
\put(28.3,9.0){\ff 2}
\put(28.3,8.2){\ff 3}
\put(27.7,8.0){\ff 1}

\put(27.7,11.2){\ff 2}
\put(28.3,11.0){\ff 0}
\put(28.3,10.2){\ff 1}
\put(27.7,10.0){\ff 3}

\put(27.7,13.2){\ff 2}
\put(28.3,13.0){\ff 0}
\put(28.3,12.2){\ff 1}
\put(27.7,12.0){\ff 3}

\put(27.7,15.2){\ff 0}
\put(28.3,15.0){\ff 2}
\put(28.3,14.2){\ff 3}
\put(27.7,14.0){\ff 1}

\end{picture}
}
\newcommand{\trellisfigfirstRSeightSTbis}[2]{
\setlength{\unitlength}{#1in}
\begin{picture}(34,16)
\put(29,15){\line(3,0){6}}
\put(29,15){\line(3,-1){6}}
\put(29,15){\line(3,-2){6}}
\put(29,15){\line(1,-1){6}}
\put(29,13){\line(1,-1){6}}
\put(29,13){\line(3,-4){6}}
\put(29,13){\line(3,-5){6}}
\put(29,13){\line(1,-2){6}}

\put(29,11){\line(3,2){6}}
\put(29,11){\line(3,1){6}}
\put(29,11){\line(3,0){6}}
\put(29,11){\line(3,-1){6}}
\put(29,9){\line(3,-1){6}}
\put(29,9){\line(3,-2){6}}
\put(29,9){\line(1,-1){6}}
\put(29,9){\line(3,-4){6}}

\put(29,7){\line(3,4){6}}
\put(29,7){\line(1,1){6}}
\put(29,7){\line(3,2){6}}
\put(29,7){\line(3,1){6}}
\put(29,5){\line(3,1){6}}
\put(29,5){\line(3,0){6}}
\put(29,5){\line(3,-1){6}}
\put(29,5){\line(3,-2){6}}

\put(29,3){\line(1,2){6}}
\put(29,3){\line(3,5){6}}
\put(29,3){\line(3,4){6}}
\put(29,3){\line(1,1){6}}
\put(29,1){\line(1,1){6}}
\put(29,1){\line(3,2){6}}
\put(29,1){\line(3,1){6}}
\put(29,1){\line(3,0){6}}
\font\ff=CMR7 
\put(0,14.5){\ff (0,2,8,10),(4,6,12,14),(5,7,13,15),(1,3,9,11)}
\put(0,12.5){\ff (20,22,28,30),(16,18,24,26),(17,19,25,27),(21,23,29,31)}
\put(0,10.5){\ff (4,6,12,14),(0,2,8,10),(1,3,9,11),(5,7,13,15)}
\put(0,8.5){\ff (16,18,24,26),(20,22,28,30),(21,23,29,31),(17,19,25,27)}

\put(0,6.5){\ff (5,7,13,15),(1,3,9,11),(0,2,8,10),(4,6,12,14)}
\put(0,4.5){\ff (17,19,25,27),(21,23,29,31),(20,22,28,30),(16,18,24,26)}
\put(0,2.5){\ff (1,3,9,11),(5,7,13,15),(4,6,12,14),(0,2,8,10)}
\put(0,0.5){\ff (21,23,29,31),(17,19,25,27),(16,18,24,26),(20,22,28,30)}
\put(29,0){\mbox{\ff \it n}}
\put(33,0){\mbox{\ff {\it n}+2}}
\font\ff=#2
\put(26.8,0.5){\ff {{7}}}
\put(26.8,2.5){\ff {{6}}}
\put(26.8,4.5){\ff {{5}}}
\put(26.8,6.5){\ff {{4}}}
\put(26.8,8.5){\ff {{3}}}
\put(26.8,10.5){\ff {{2}}}
\put(26.8,12.50){\ff {{1}}}
\put(26.8,14.5){\ff {{0}}}
\put(27.5,0){\line(0,1){16}}
\font\ff=CMR7 
\put(27.7,1.2){\ff 3}
\put(28.3,1.0){\ff 1}
\put(28.3,0.3){\ff 0}
\put(27.7,0.0){\ff 2}

\put(27.7,3.2){\ff 1}
\put(28.3,3.0){\ff 3}
\put(28.3,2.2){\ff 2}
\put(27.7,2.0){\ff 0}

\put(27.7,5.2){\ff 1}
\put(28.3,5.0){\ff 3}
\put(28.3,4.2){\ff 2}
\put(27.7,4.0){\ff 0}

\put(27.7,7.2){\ff 3}
\put(28.3,7.0){\ff 1}
\put(28.3,6.2){\ff 0}
\put(27.7,6.0){\ff 2}

\put(27.7,9.2){\ff 0}
\put(28.3,9.0){\ff 2}
\put(28.3,8.2){\ff 3}
\put(27.7,8.0){\ff 1}

\put(27.7,11.2){\ff 2}
\put(28.3,11.0){\ff 0}
\put(28.3,10.2){\ff 1}
\put(27.7,10.0){\ff 3}

\put(27.7,13.2){\ff 2}
\put(28.3,13.0){\ff 0}
\put(28.3,12.2){\ff 1}
\put(27.7,12.0){\ff 3}

\put(27.7,15.2){\ff 0}
\put(28.3,15.0){\ff 2}
\put(28.3,14.2){\ff 3}
\put(27.7,14.0){\ff 1}

\end{picture}
}

\title{Fading-Resilient
Super-Orthogonal Space-Time Signal Sets: Can Good Constellations Survive in  Fading?}

\author{D.\ Mihai Ionescu,~\IEEEmembership{Senior Member,~IEEE,} and Zhiyuan~Yan,~\IEEEmembership{Member,~IEEE} %
\thanks{D.\ Mihai Ionescu is with Nokia Research Center, 6000
Connection Drive, Irving, Texas 75039 USA (e-mail:
michael.ionescu@nokia.com).}
\thanks{Zhiyuan Yan was with the Department of Electrical and
Computer Engineering, University of Illinois at Urbana-Champaign.
He is now with the Department of Electrical and Computer
Engineering, Lehigh University, 19 Memorial Drive West, Bethlehem,
PA 18015 USA (e-mail: yan@lehigh.edu).}%
}

\date{}
\markboth{Submitted to IEEE Transactions on Information Theory}{Ionescu and Yan: Fading-Resilient Multidimensional Space-Time Constellations}

\maketitle
\begin{abstract}
In this correspondence, first-tier indirect (direct) discernible constellation expansions are defined for generalized orthogonal designs. The expanded signal constellation, leading to so-called super-orthogonal codes, allows the achievement of coding gains in addition to diversity gains enabled by  orthogonal designs. Conditions that allow the shape of an expanded  multidimensional constellation to be preserved at the channel output, on an instantaneous basis, are derived. It is further shown that, for such constellations, the channel alters neither the relative distances nor the  angles between  signal points in the expanded signal constellation.
\end{abstract}
\begin{keywords}
Euclidean distance, fading channels, geometrical uniformity,
space-time codes, constellation space invariance, fading resilience.
\end{keywords}

\section{Introduction}
\font \hollow=msbm10 at 12pt 
Encoding jointly along spatial and temporal dimensions has
received considerable attention over the recent years, and the
concerted research effort has led to improved understanding of
both block and trellis designs of space-time codes (see, for
example,
\cite{special,ala:sim,tar:spa,tar:ort,tir:squ,IonescuMYL:01,ses:sup,jaf:sup,siw:impcon,siw:import,big:per,LWKC:03,Ionescu:03});
by comparison, geometric considerations have been sporadic (see,
e.g., \cite{sch:geo,yan:geo})---perhaps due to the perception that
multiplicative distortions incurred as a result of fading can
destroy symmetries. Contrary to any such perception, Schulze
proved in \cite{sch:geo}
that flat fading channels leave invariant the shape of (generalized) orthogonal space-time constellations (or codematrices)---although he viewed his results mainly as a geometrical interpretation for the optimal detection of all orthogonal and generalized orthogonal space-time constellations, which are linearly decodable; note that generalized orthogonal designs are alternatively called space-time block codes.
Recently, Gharavi-Alkhansari and Gershman \cite{gha:con} examined the same invariance property for an orthogonal space-time constellation, and used it to explain why optimal decoding  reduces to symbol-by-symbol decoding.

There exists an alternative motivation for examining the
conditions that allow the shape of a constellation to be preserved
in fading; it pertains to code designs that rely on certain
geometrical properties of the (multidimensional) constellation,
such as the spectrum of relative Euclidean distances. When
performance is viewed on an instantaneous basis, rather than on
average, the observed  relative distance between two valid points
(codewords) depends on the effect of multiplicative distortion
(fading) on the two points;
if an instantaneous
realization of the channel distorts  valid candidate points differently,
a less likely point may appear more likely at the channel output,
with respect to receiver observations\footnote{When performance,
e.g.\ pairwise error probability, is averaged over  channel
realizations the result  only depends  on constellation
properties, in isolation of fading;  this makes the effect of
instantaneous channel distortions transparent to performance {\em
on average}.}. When instantaneous performance---as a function of
Euclidean distances---is relevant, it becomes crucial to be able
to preserve the shape of the signal constellation for any
realization of the channel (multiplicative distortion). This is
the motivation for considering the resilience of the constellation
shape to fading; other implications of fading resilience are
discussed below.

In \cite{sch:geo}, a necessary and sufficient condition \cite{tir:squ} for orthogonality of a space-time constellation---such as arising, e.g., from  Radon-Hurwitz constructions \cite{ala:sim}---shows that when a detector operates to detect individual coordinates\footnote{That is, it provides information on real coordinates of complex symbols that make up the space-time constellation, rather than on the complex entities themselves; this is always the case.}, the detection equation at any receive antenna is such that the equivalent channel leaves invariant---up to a scaling factor---the distances between the (potentially transmitted) multidimensional space-time constellation points, as well as 
their
respective angles;
one can recognize
this invariance
to be a form of resilience to fading of the (generalized)
orthogonal space-time constellation, whose shape is, in effect,
preserved (up to a scaling factor) in spite of the multiplicative
distortions due to flat fading. Note that the above assumption
about coordinate-wise detection implies that a multidimensional
space-time constellation point from ${\ho C}^{n_0}$, $n_0 \in {\ho
N}$, is viewed (by the detector) as a point from ${\ho
R}^{2{n_0}}$, via a well-known isometric transformation (see,
e.g., \cite[eq.\ (1)]{sch:geo}). As demonstrated in
\cite{sch:geo}, the invariance property applies directly to
space-time codematrices from (generalized) orthogonal designs
\cite{tar:ort} or from unitary designs \cite{tir:squ}---mainly
because such designs allow any space-time constellation point to
be expressed as a linear combination of basis matrices (see proof
in \cite{sch:geo}). It is known that even an orthogonal space-time
block code that has full-rate is, in essence, a space-time
modulator; i.e., it can provide diversity gain in flat fading
channels, but no coding gain (as redundancy is inserted in the
spatial dimension, and the inherent repetition in the time
dimension provides as good a coding redundancy as repetition codes
do)\footnote{E.g., when simulating in AWGN channels an Alamouti
code \cite{ala:sim} with two transmit and one receive antennas,
whereby the complex elements mapped to $2 \times 2$ space-time
complex matrices are drawn from a 4PSK constellation, one obtains
the familiar uncoded performance of 4PSK in AWGN---provided that
both schemes use the same total energy per channel use, or
equivalently the (average) received bit energy values per receive
antenna are the same.}. Linearly decodable, real,
generalized orthogonal designs (respectively complex unitary designs)
for $N$ transmit antennas can be viewed as $T \times
N$ real matrices (respectively $N \times N$ complex matrices);
they are (non-surjective) mappings from ${\ho R}^{2K}$ to ${\ho
R}^{TN}$ (respectively to ${\ho R}^{2N^2}$), where $T$ is the number
of channel uses (symbol epochs) covered by a codematrix
\cite[Definition 4]{tir:squ}.
In light of the need to add coding gain,
the natural question is whether this fading resilience can be preserved when coding redundancy is added---preferrably, without modifying the spectral efficiency or expanding the bandwidth; this, in turn, requires that the space-time constellation be extended beyond the orthogonal set. Such constructions have been reported by Ionescu {\em et al.} \cite{IonescuMYL:01}, and later generalized by Siwamogsatham and Fitz \cite{siw:impcon,siw:import} and by Seshadri and Jafarkhani \cite{ses:sup,jaf:sup}, who dubbed such codes `super-orthogonal'.
It is shown in this correspondence that Schulze's result for (generalized) orthogonal designs can be extended 
to a larger family of space-time constellations, and 
to space-time codes that are not linearly decodable. 
If, in addition, the space-time constellation and the redundancy
scheme itself have additional symmetries related to the shape of
the constellation and the codebook---e.g., geometrical uniformity
\cite{yan:geo}---then the important implication would be  that
symmetries such as geometrical uniformity {\em can} be preserved
after passing through the fading channel. This, in turn, should
motivate efforts to embed symmetry enabling structures into
codes designed
for fading channels. In particular, fading resilient symmetries
are enablers for extending the concept of 
geometrical uniform codes to multiple-input-multiple-output (MIMO)
fading channels---e.g., by using such constellations along with a
more powerful redundancy scheme, such as a turbo, multilevel, or
low-density parity-check (LDPC) code.

\section{Fading Resilience via Geometrically Invariant Properties}
Let $i=\sqrt{-1}$ and consider a linearly decodable, complex,
linear, generalized orthogonal design $\cal O$ of rate $K/T$ for
$N$ transmit antennas, which maps a vector $\sw = [z_1, \ldots ,
z_K]^{\rm T} \in {\ho C}^{K}$ of $K$ complex symbols $z_k \df x_k
+ i y_k$, $k=0, \ldots , K-1$, to  semiunitary complex $T \times
N$ matrices $\SM \in {\cal M}_{T,N}({\ho C})$; semiunitarity means
that $\SM^{\rm H} \SM = \|\sw\|^2 \bd{I}_N$ (even when $T \neq
N$), and the linearly decodable assumption leads, in one aspect,
to the constraint $T\geq N$. The constraint $T\geq N$ can be
dropped if one considers codes that are not linearly decodable. In
another aspect, pursuant to the isometry ${\cal I} : {\ho C}^K
\mapsto {\ho R}^{2 K}$ that maps $\sw$ to the $2K$-dimensional
real vector $\chiw \df [\Re \{z_1\}, \Im \{z_1\}, \ldots, \Re
\{z_K\}, \Im \{z_K\}]^{\rm T}={\cal I}(\sw)$, linearity in the
arbitrary symbols $z_k$, $k=0, \ldots , K-1$, means that there
exist $2K$ basis matrices of size $T \times N$, with complex
elements, such that
\begin{eqnarray}
\lefteqn \SM & = &\!\! \sum_{l=0}^{2K-1} \chi_l \betaM_{l} \in {\cal O}, \ \ \ \forall \bd{\chi} \in {\ho R}^{2 K}  \label{linexp}\\
 & = & \!\!\!\!\! \sum_{l=1}^{K} \left( x_l \betaM_{2l-2} + y_l \betaM_{2l-1}\right)
=\!\! \sum_{l=1}^{K} \left( z_l \betaM^-_{l} + z^\ast_l
\betaM^+_{l}\right),
\label{linexpbis}
\end{eqnarray}
where the asterisk represents complex conjugation, and
\cite{tir:squ} \be \betaM^\pm_l=\frac{1}{2}\left(\betaM_{2l-2} \pm i
\betaM_{2 l -1}  \right); \label{betapm} \ee
a necessary and sufficient condition for $\SM^{\rm H} \SM = \|\sw \|^2 \bd{I}_N$ is  
\be
{\textstyle
\betaM_l^{\rm H} \betaM_p + \betaM_p^{\rm H} \betaM_l = 2 \delta_{l p} \bd{I}_N, \ l,p =0, \ldots, 2K-1
},
\label{nsc}
\ee
where $\bd{I}_N$ is the $N \times N$ identity matrix.
\newtheorem{rem1}{Remark}
\begin{rem1}
The rate $K/T$ mentioned above represents only a symbol rate, which does not indicate in any way a (finite) spectral efficiency---unless the complex symbols are restricted to a common finite constellation $\cal Q$ such as $m$-PSK, with $m$ some integer power of 2; in other words, the complex symbols $z_k$'s (or the real $2K$-tuple $\chiw$) can assume arbitrary complex (real)  values
($\cal O$ is non-countable).
\end{rem1}

As long as $\chiw \in {\ho R}^{2 K}$, the set $\cal O$ spanned by
the basis $\{\betaM_l\}_{l=0}^{2 K-1}$ over $\ho R$ is a vector
space. Specifying a (finite) spectral efficiency means, e.g.,
restricting the complex symbols $z_k$, $k=0, \ldots , K-1$, to a
common finite constellation $\cal Q$, e.g.\ $m$-PSK; this will
produce a multidimensional space-time constellation with a finite
cardinality, denoted $\cal G \subset \cal O$ in the sequel;
nevertheless, eqs.\ \rf{linexp} and \rf{nsc} still hold because
$\cal Q \subset {\ho C}$ and, respectively, because restricting
$z_k$, $k=0, \ldots , K-1$, to $\cal Q$ does not modify the basis
expansion in $\cal O$. Note that \rf{linexp}, \rf{nsc} directly
lead to \be {\textstyle (\SM -\SM')^{\rm H}(\SM -\SM') = \|\chiw
-\chiw'\|^2 \bd{I}_N, \ \ \ \forall \SM, \SM' \in {\cal O} }.
\label{eec} \ee Since the complex Radon-Hurwitz eqs.\ \rf{nsc} are
invariant to multiplication of all matrices in a generator set by
$\zeta \in {\ho C}, |\zeta|=1$, it follows that
$\{\betaM_l\}_{l=0}^{2 K-1}$ is a basis in $\cal O$ if and only if
$\{\betaM_l \zeta \}_{l=0}^{2 K-1}$ is.

An expansion (see below) of the finite space-time constellation $\cal G$---as practiced, e.g., in \cite{IonescuMYL:01,siw:impcon,siw:import,ses:sup,jaf:sup}---does not necessarily
remain {\em within} the limits of the generalized orthogonal design $\cal O$, and orthogonality of pairwise differences [see \rf{eec}] is not necessarily preserved in the expanded constellation.

\subsection{Constellation Expansions and Their Properties}
As mentioned above, adding coding redundancy without modifying the
spectral efficiency requires that the finite space-time
constellation be extended beyond the set $\cal G$ of orthogonal
matrices. Consider a multidimensional space-time constellation
$\cal G$ from a generalized complex orthogonal design $\cal O$,
and an expansion of $\cal G$ via a
symmetry or
by multiplication with
some unitary $N \times N$ matrix $\bd{U}$. 
A first-tier expanded constellation is
\be
{\cal G}_{\rm e} \df {\cal G} \cup {\cal G} \bd{U}.
\label{ge}
\ee
and has been introduced in \cite{IonescuMYL:01}. 
Specifically,
with a 4PSK constellation on each of $N=2$ transmit antennas, \cite{IonescuMYL:01} used a symmetry operation
(characterized further in
\cite[Section II.B]{yan:geo})
 to expand an orthogonal set of sixteen matrices obtained by mapping all $K$-tuples of 4PSK elements to $T \times N$ matrices, where $K=T=2$; after expansion, pairwise differences are in general non-orthogonal (no longer verify \rf{eec}), and the
symmetry operation used in \cite{IonescuMYL:01} corresponds to right multiplication 
by the unitary matrix $\mtrx{1}{0}{0}{-1}$---recognized to be
a particular case of the `super-orthogonal' construction from \cite{ses:sup,jaf:sup}. Note that any symmetry can be described as multiplication by a unitary matrix of appropriate size.
\newtheorem{rem2}[rem1]{Remark}
\begin{rem2}
It should be stressed here that, whenever the intention is to guarantee some geometrical invariance property of the expanded constellation ${\cal G}_{\rm e}$, the preferred method for expanding $\cal G$ should be some symmetry operation, rather than an arbitrary unitary transformation---which, in turn, should arise simply as a consequence of the symmetry itself; the reason is, of course, the very nature of the expected result, which is some form of geometrical invariance.
\label{rem2}
\end{rem2}

As already noted,
${\cal G} \bd{U} \not\subset {\cal O}$, in general, because ${\cal G} \bd{U}$ is not necessarily in the span of $\{\betaM_l\}_{l=0}^{2 K-1}$; thereby, orthogonality of pairwise differences after a constellation expansion that does not alter the spectral efficiency will be lost.
Nevertheless, if $\SM \in {\cal G}$, then $(\SM \bd{U})^{\rm H} \SM \bd{U} = \|\sw\|^2 \bd{I}_N$ 
and 
\begin{eqnarray}
\SM \bd{U}&=& \mbox{$\sum_{l=0}^{2K-1} \chi_l \betaM'_l$}, \ \ \ \forall \chiw \in {\ho R}^{2 K} \label{linexpp}\\
\betaM'_l &=&  \betaM_l \bd{U}, \ \ \ \forall l=0, \ldots, 2 K-1 \label{newbetas}
\end{eqnarray}
As discussed above \cite{tir:squ}, $\{\betaM_l\}_{l=0}^{2 K-1}$
verify the complex Radon-Hurwitz eqs.\ \rf{nsc}, while $\{\betaM'_l\}_{l=0}^{2 K-1}$ verify
\be
{
{\betaM'_{l}}^{\rm H} \betaM'_p + {\betaM'_{p}}^{\rm H} \betaM'_l = 2 \delta_{l p} \bd{I}_N, \ l,p =0, \ldots, 2K -1
};
\label{nscp}
\ee
however, a similar property does
not necessarily hold for two basis matrices from the different
sets $\{\betaM_l\}_{l=0}^{2 K-1}$, $\{\betaM'_l\}_{l=0}^{2 K-1}$.

Since $\bd{U}$ is unitary if and only if $\bd{U}\zeta$ is unitary---provided that $\zeta \in {\ho C}, |\zeta|=1$---expansions via $\bd{U}\zeta$ and $\bd{U}$ should be simultaneously characterizable as applying  $\bd{U}$ to either $\cal G \zeta$ or $\cal G$.

\newtheorem{l0}[rem1]{Lemma}
\begin{l0}
\label{l0}
Let ${\cal Q} \in {\ho C}$ be a (finite) complex constellation, and  $\{\betaM_l\}_{l=0}^{2 K-1}$  a generator set for $\cal G \subset {\cal O}$ over ${\cal I}({\cal Q}^{K})$, such that any $\SM \in {\cal G}$ verifies \rf{linexp}, \rf{linexpbis} with $z_k \in {\cal Q}$. Let $\zeta \in {\ho C}, |\zeta|=1$. Then
\begin{eqnarray}
\lefteqn{\tilde{\protect{\cal G}}  \df {\cal G} \zeta  =  \left\{ \tilde{\SM} \df \SM \zeta | \SM \in {\cal G} \right\} \subset {\cal O},} \!\!\!\! \\
\tilde{\SM} & = & \mbox{$\sum_{l=1}^{K} \left[ \Re\{z_l \zeta\} \eta_{2 l -2} + \Im\{z_l \zeta\} \eta_{2 l -1} \right]$}, \label{linexprot}\\
\!\!\!\tilde{\SM} \!\!\!& \!\!= \!\!&\!\!\!\!\!\! \mbox{$\sum_{l=1}^{K} \left[ \tilde{x}_l  \eta_{2 l -2} + \tilde{y}_l  \eta_{2 l -1} \right]$}, \ \tilde{x}_l + i \tilde{y}_l \df \tilde{z}_l \in {\cal Q}\zeta, \\
\!\! \eta_{2 l -2} \!\! \!\! & \!\! \df \!\! \!\! &\!\! \zeta \left(\Re\{\zeta\} \betaM_{2 l -2} \! - \Im\{\zeta\} \betaM_{2 l -1}\right), l=1, \ldots , \! K, \label{etae}\\
\!\! \eta_{2 l -1} \!\! \!\! & \!\! \df \!\! \!\! &\!\! \zeta \left(\Re\{\zeta\} \betaM_{2 l -1} \! + \Im\{\zeta\} \betaM_{2 l -2}\right), l=1, \ldots , \! K. \label{etao}
\end{eqnarray}
Moreover, $\{\etaM_l\}_{l=0}^{2 K-1} \subset {\cal O}$ and 
\be
{\textstyle
\etaM_l^{\rm H} \etaM_p + \etaM_p^{\rm H} \etaM_l = 2 \delta_{l p} \bd{I}_N, \ l,p =0, \ldots, 2K-1
}.
\label{nsce}
\ee
\end{l0}

\begin{proof}
A sketch of proof is as follows. The fact that $\{\etaM_l\}_{l=0}^{2 K-1} \subset {\cal O}$ is obvious; simple manipulations of \rf{etae}, \rf{etao}, \rf{nsc} prove  \rf{nsce} directly. To prove \rf{linexprot} it suffices to re-write  the terms in the second summation of \rf{linexpbis} as $z_l \betaM^-_{l} + z^\ast_l \betaM^+_{l} = z_l \zeta \zeta^{\ast} \betaM^-_{l} + z^\ast_l \zeta^{\ast}\zeta  \betaM^+_{l}= (z_l \zeta) \etaM^-_{l} + (z_l \zeta)^{\ast}\etaM^+_{l}$, where $\etaM^+_{l} = \betaM^+_{l} \zeta$ and $\etaM^-_{l} = \betaM^-_{l} \zeta$, followed by straightforward manipulations and by finally multiplying \rf{linexpbis} by $\zeta$.
\end{proof}

Lemma~\ref{l0} shows that an expansion of $\cal G$ by ${\cal G} \zeta = {\cal G} (\zeta \bd{I}_N)$ simply changes the generator set and the alphabet (from $\cal Q$ to ${\cal Q} \zeta$), and is {\em indiscernible} (from $\cal O$) in the sense that ${\cal G} \zeta \subset {\cal O}$. Therefore {\em expansions of the form ${\cal G}_{\rm e} = {\cal G} \cup {\cal G} \bd{U} \zeta$ differ from those of the form ${\cal G}_{\rm e} = {\cal G} \cup {\cal G} \bd{U}$ only in that $\bd{U}$ operates on a different subset of $\cal O$} (${\cal G}\zeta$ vs.\ ${\cal G} $). Clearly, $\zeta \in {\ho C}, |\zeta|=1$  preserves the constellation energy.

\newtheorem{rem3}[rem1]{Definition}
\begin{rem3}
If  $\zeta \in {\ho C}, |\zeta|=1$, $\zeta \not = 1$, and $\bd{U}\not= \bd{I}_N$ is a $N \times N$ unitary matrix, then a first-tier, indirect (direct), discernible constellation expansion of $\cal G$ is
${\cal G}_{\rm e} = {\cal G} \cup {\cal G} \bd{U} \zeta$ \ \ (${\cal G}_{\rm e} = {\cal G} \cup {\cal G} \bd{U}$), where ${\cal G} \bd{U} \zeta \not = {\cal G} $ (${\cal G} \bd{U} \not = {\cal G}$)
and
$\bd{U}$ has either more than two distinct eigenvalues, or all
real eigenvalues.\footnote{Via Lemma~\protect\ref{l0}, this
accommodates constellation expansions by a unitary (not
necessarily Hermitian) matrix that has complex eigenvalues, but
only arising as a rotation of a set of real eigenvalues.}
\end{rem3}

Consider a direct discernible constellation expansion of $\cal G$ to ${\cal G} \cup {\cal G} \bd{U}$, where matrices $\SM$, $\SM \bd{U}$ verify \rf{linexp}, \rf{linexpp} $\forall \SM \in {\cal G}$.

\newtheorem{l1}[rem1]{Theorem}
\begin{l1}
\label{l1}
If ${\cal G}_{\rm e}$ of \rf{ge} is  a first-tier, direct, discernible expansion by $\bd{U}\not=\pm \bd{I}_N$ of a multidimensional space-time constellation $\cal G$ from a generalized complex orthogonal design, having a generator set $\{\betaM_l\}_{l=0}^{2 K-1}$, and if $\{\betaM'_l\}_{l=0}^{2 K-1}$ is the generator set for ${\cal G}' \df {\cal G} \bd{U}$ that verifies \rf{newbetas}, then no element of the set $\{\betaM'_l\}_{l=0}^{2 K-1}$ is  a linear combination, over $\ho R$, of the matrices $\betaM_l$, $l=0, \ldots , 2 K-1$. 
\end{l1}

\begin{proof}
Assume to the contrary that $ \betaM'_{q_0} = \betaM_{q_0} \bd{U}
= \sum_{q=0}^{2 K-1} t_q \betaM_q$, where $\bd{t} \df [t_0, \ldots
, t_{2 K-1}]^{\rm T} \in {\ho R}^{2 K}$.  It can be easily
verified, using \rf{nscp}, that $\sum_{q=0}^{2 K-1} t_q^2 =1$.
First, assume that at least two components of $\bd{t}$ are
nonzero. Then, for some nonzero $t_{q_1}$, $q_1 \not = q_0$,
$\betaM_{q_1}= t_{q_1}^{-1} \betaM_{q_0} \bd{U} - t_{q_1}^{-1}
\sum_{q\not = q_1, q_0} t_q \betaM_q - t_{q_0}t_{q_1}^{-1}
\betaM_{q_0}$. From \rf{nsc}, $\betaM_{q_1}^{\rm H} \betaM_{q_0} +
\betaM_{q_0}^{\rm H} \betaM_{q_1} = 0$, which can be reduced after
straightforward manipulations to $t_{q_1}^{-1} \bd{U}^{\rm H}
\betaM_{q_0}^{\rm H} \betaM_{q_0} - t_{q_1}^{-1} \sum_{q\not =
q_1, q_0} t_q (\betaM_{q}^{\rm H} \betaM_{q_0} + \betaM_{q_0}^{\rm
H} \betaM_q) - 2 t_{q_0}t_{q_1}^{-1}\betaM_{q_0}^{\rm H}
\betaM_{q_0} + t_{q_1}^{-1} \betaM_{q_0}^{\rm H} \betaM_{q_0}
\bd{U} =0 $, or, after using \rf{nsc}, $t_{q_1}^{-1} \bd{U}^{\rm
H}  - 2 t_{q_0}t_{q_1}^{-1} \bd{I}  + t_{q_1}^{-1} \bd{U} =0 $.
Then $\bd{U}^{\rm H}= 2 t_{q_0} \bd{I} -\bd{U}$, and unitarity of
$\bd{U}$ translates into $\bd{U}$ verifying the equation \be
\bd{U}^2 - 2 t_{q_0} \bd{U} + \bd{I} =0. \label{me} \ee
Assume that $\bd{U}$ verifies a (monic) polynomial equation of degree smaller than two, namely $\bd{U} + m_0 \bd{I}=0$; then, $\bd{U}=-m_0 \bd{I}$, and unitarity together with the assumption that $\bd{U}$ has real eigenvalues imply that $\bd{U} =\pm \bd{I}_N$, which contradicts the hypothesis. Then, necessarily, \rf{me} is the minimum equation of $\bd{U}$. But $t^2 -2 t_{q_0} t +1 =0$ has roots $t^{(1),(2)}=t_{q_0} \pm \sqrt{t^2_{q_0}-1}$, with $t_{q_0}<1$; thereby, since the irreducible (in $\ho C$, in this case) factors of the minimum polynomial divide the characteristic polynomial, it follows that the distinct eigenvalues of $\bd{U}$ are the distinct roots  among $\{t^{(1)}, t^{(2)}\}$, which do have, indeed, unit magnitude, but nonzero imaginary parts---again contradicting the hypothesis. Finally, assume that  only one component of $\bd{t}$ is nonzero, say $\betaM'_{q_0} = \betaM_{q_0} \bd{U} =\betaM_{q_1}$, $q_1 \not = q_0$. Then \rf{nsc} 
is equivalent to $\bd{U}^{\rm H} + \bd{U}=0 \Leftrightarrow \bd{U}^2 + \bd{I} =0$, and the minimal polynomial $t^2 +1 =0$ has non-real  roots $\pm i$---again contradicting the hypothesis.
This completes the proof.
\end{proof}

Since ${\cal G} \zeta \subset {\cal O}$, as discussed above, a
similar contradiction as the one used above can be employed to
infer directly
\newtheorem{l3}[rem1]{Corollary}
\begin{l3}
\label{l3}
If ${\cal G}_{\rm e} = {\cal G} \cup {\cal G} \bd{U} \zeta$ is a discernible expansion then
$({\cal G}_e \backslash {\cal G} ) \cap {\cal O} =\{0\}$.
\end{l3}

Thereby,  Theorem~\ref{l1} leads directly to a direct sum structure via
\newtheorem{l2}[rem1]{Corollary}
\begin{l2}
\label{l2}
Any discernible expanded constellation ${\cal G}_{\rm e}$ is naturally embedded in a direct sum of two $2 K$-dimensional vector sub-spaces of ${\cal M}_{T,N}({\ho C})$, and
\be
{\textstyle
\SM = \sum_{l=0}^{2 K-1} \chi_l \betaM_l + \sum_{l=0}^{2 K-1} \chi'_l \betaM'_l, \ \ \ \forall \SM \in {\cal G}_{\rm e}.
}
\label{dirprod}
\ee
\end{l2}

\subsection{Implications of Discernible Constellation Expansions}
In all cases where the Euclidean distance between points from  the
multidimensional constellation ${\cal G}_{\rm e}$ is relevant
\cite{LWKC:03,big:per,Ionescu:03,yan:geo}, the Euclidean, or
Frobenius, norm of $\SM \in {\cal G}_{\rm e}$ is important; then,
$\SM$ can be identified via an isometry with a vector from ${\ho
R}^{2 T N}$, where $2 T N$ is the total number of real coordinates
in $\SM$ 
when using the expanded constellation
${\cal G}_{\rm e}$. Therefore, since $\SM \in {\cal G}_{\rm e}$ is
completely described by the $2 \cdot 2 \cdot K$ real coordinates
of the embedding space (see \rf{dirprod}), it follows that the
first tier expansion uses $4 K$ of the available $2 T N$ diversity
degrees of freedom. Note that, since when $N\geq 2$ the maximum
rate for square matrix embeddable space-time block codes (unitary
designs) is at most one \cite[Theorem 1]{tir:squ}, it follows that
$K\leq T$ and the dimensionality condition implicit in
\rf{dirprod} is well-defined.

\subsection{Fading Resilience}
In order to show that ${\cal G}_{\rm e} = {\cal G} \cup {\cal G} \bd{U}$ is resilient
to flat fading, assume that a  code matrix $\bd{c} \in {\cal
G}_{\rm e}$, is selected for transmission  from the $N$ transmit antennas
during $T$ time epochs
; an arbitrary element of ${\cal G}_{\rm e}$ (denoted $\bd{S}$ in above paragraphs) verifies \rf{dirprod}, and either the $\chi_k$ coefficients  or the $\chi'_k$ coefficients vanish.
Without loss of generality, assume there is one receive antenna.
Clearly, the code matrix selected for transmission verifies either   $\bd{c} \in {\cal G}$ or $\bd{c} \in {\cal G}_{\rm e} \backslash  {\cal G} $; assume first the former,
i.e.\ all $\chi'_k$ coefficients vanish in \rf{dirprod}. The observation vector 
during the $T$ time epochs is given by
\[\bd{r}=\bd{c}\bd{h}+\bd{n_c},\]
where $\bd{h}=[h_1\,\,h_2\,\,\cdots h_N]^T$ is the vector of
complex multiplicative fading coefficients and $\bd{n_c}$ is
complex AWGN with variance $\sigma ^2=N_0/2$ in each real
dimension. Given $\bd{h}$ and $\bd{n_c}$, when $\chi'_k$'s are
all zeros, the received vector is simply
\[{\textstyle \bd{r}=\sum _{k=0}^{2K-1}{\chi_k \etaM _k}+\bd{n_c}},\]
where $\etaM _k=\betaM _k \bd{h}$ for $k=0$, $1$, $\cdots$,
$2K-1$. By eq.\ (\ref{nsc}), it can be shown that ${\Re}\{ \langle
\etaM_k, \etaM_l \rangle \}= \| \bd{h} \|^2 \delta_{k l}$. Define
$\bd{g}_k$ as the real vector corresponding to $\etaM_k$ as
follows:
\[\etaM_k \leftrightarrow \| \bd{h} \| \bd{g}_k  \mbox{ for } k=0, 1, \cdots, 2K-1,\]
where $\leftrightarrow$ denotes the correspondence between complex
and real vectors.  Clearly, $\bd{g}_k$'s are real orthonormal
vectors. Also define the real vectors corresponding to $\bd{r}$
and $\bd{n_c}$ respectively as follows: $\bd{r} \leftrightarrow
\bd{y}$ and $\bd{n_c} \leftrightarrow \bd{n}$. Then, the received
real vector
\[{\textstyle \bd{y}=\| \bd{h} \| \sum _{k=0}^{2K-1}{\chi_k \bd{g}_k}+\bd{n}.}\]
Define $\bd{G}=[\bd{g}_0\ \bd{g}_1\, \cdots \, \bd{g}_{2K-1}]$,
$\bd{\chi}=[\chi_0\, \cdots \, \chi_{2K-1}]^T$; then
\[\bd{y}=\|\bd{h}\| \bd{G}\bd{\chi}+\bd{n}.\]
 Similarly, when $\bd{c} \in {\cal G}_{\rm e} \backslash  {\cal G} $, i.e.\ all $\chi_k$'s in \rf{dirprod} vanish, the following equation holds:
\[{\textstyle \bd{y'}=\| \bd{h} \| \sum _{k=0}^{2 K -1}{\chi_k' \bd{g}_k'}+\bd{n'}},\]
where $\bd{r}\leftrightarrow \bd{y}'$ and $\betaM _k' \bd{h}
\leftrightarrow \| \bd{h} \| \bd{g}_k'$. That is,
\[\bd{y}'=\|\bd{h}\| \bd{G}' \bd{\chi}'+\bd{n}',\]
where $\bd{G}'=[\bd{g}'_0\, \bd{g}'_1\, \cdots\, \bd{g}'_{2K-1}]$,
$\bd{\chi}'=[\chi'_0\, \chi'_1\, \cdots \, \chi'_{2K-1}]^T$.
Conditioned on whether the transmitted signal point is selected
from ${\cal G}$ or from ${\cal G}_{\rm e} \backslash  {\cal G}$,
one  can first define $\bd{\chi}_{\oplus}=[\bd{\chi}^T\,
\bd{\chi}'^T]^T$, $\bd{n}_\oplus=[\bd{n}^T\, \bd{n}'^T]^T$, and
$\bd{y}_\oplus =[\bd{y}^T\, \bd{y}'^T]^T$, where either half of
the real coefficients vanish,
 then express the
received signal in both cases as \be \bd{y}_\oplus=\|\bd{h}\|
\bd{G}_\oplus \bd{\chi}_\oplus+\bd{n}_\oplus
\label{anglepreserving}\ee where $\bd{G}_\oplus$ is the $2\cdot 2
\cdot T \times 2\cdot 2 \cdot K $ matrix $\left[\begin{array}{cc}
\bd{G} & 0\nonumber \nonumber \\
0 & \bd{G}' \nonumber
\end{array} \right]$. It is easy
to verify that $\bd{G}_\oplus^{T} \bd{G}_\oplus=\bd{I}_{2 \cdot 2
\cdot K }$. Hence, $\bd{y}_\oplus$ preserves the distances and
angles of $\bd{\chi}_\oplus$---up to the scaling factor
$\|\bd{h}\|$ and noise.

A final discussion pertains to the side information on whether the transmitted signal point belongs to
${\cal G}$ or 
${\cal G}_{\rm e} \backslash  {\cal G}$:
\begin{enumerate}
\item Representing the multidimensional points in ${\cal G}_{\rm
e}$---and their respective Euclidean distances---in terms of
vectors coordinates ($\chi_k$, $\chi'_k$) rather than matrix
entries, was preferred above only because it simplified the
analysis; \item The  side information mentioned above is naturally
available at the receiver during hypothesis testing---since any
tested point in ${\cal G}_{\rm e}$ belongs to an unique
subconstellation, thereby allowing one to form $\bd{\chi}_\oplus$
by appropriate zero-padding; then, for each hypothesis, the
nonzero received (i.e., observed) coordinates can be easily padded
with leading or trailing zeroes, in order to form $\bd{y}_\oplus$
and match the standing hypothesis about the transmitted point.
Thereby, when testing various $\bd{\chi}_\oplus$ vectors---from a
constellation ${\cal G}_{\rm e}$ with a given shape---performance
is determined precisely by the distances and angles between
$\bd{y}_\oplus$ vectors; if the latter match the distances and
angles between points in ${\cal G}_{\rm e}$ (up to noise, and a
scaling factor due to fading), then the shape of  ${\cal G}_{\rm
e}$ is preserved, and other symmetry properties of ${\cal G}_{\rm
e}$ become relevant when they exist. \item Equivalently, rather than calculating the Euclidean distances
between multidimensional points from ${\cal G}_{\rm e}$ 
in terms of {\em vector} coordinates $\chi_k$, $\chi'_k$, the
decoder may (and usually does) compute them as Frobenius norms of
(respective difference) {\em  matrices}. (Euclidean distances
between $\bd{\chi}_\oplus$ vectors and Frobenius norms of their
corresponding difference matrices are the same---with proper
normalization.) \item For example, in the space-time trellis codes
from \cite{IonescuMYL:01}\footnote{Same appears to be true of  codes ({\em
not} extended constellation!) from \cite{jaf:sup}.  
}, the branches departing from, and converging to, any state
use signal points from one subconstellation; when a maximum
likelihood receiver tests any branch, the originating state of the
branch together with the associated information bits determine a
point from a precise subconstellation.
\end{enumerate}

Hence, the decoder on the
receiver side does, naturally, have access to the 
side information during hypothesis testing, 
and thereby benefits from shape invariance.

In summary, the fading channel, up to scaling and noise, leaves
invariant the shape in the expanded signal constellation ${\cal
G}_{\rm e}$. Although the maximum likelihood decoding for the
expanded signal constellation is no longer linear, the decoding
process benefits from this property nonetheless.

\section{Example}
 \tbbBIS{tbh} \tbQPSK{tbh} In this section, we illustrate the above results with
the expanded signal constellation in \cite{IonescuMYL:01}.

The expanded signal constellation in \cite{IonescuMYL:01} over
QPSK is shown in Table~\ref{tab:superc}. The entries in the
codematrices in Table~\ref{tab:superc} are the indices of the
signal points in Table~\ref{tab:qpsk}. It is clear that the first
$16$ matrices, ${\mathcal C}_i$ ($0\le i \le 15$), are of the form
$\left[\begin{matrix}
A & B^{\ast} \\
B& -A^{\ast}
\end{matrix} \right]$, and hence can be expressed as linear combinations of the following four base
matrices:
\[
\frac{1}{\sqrt{2}}\mtrx{1}{0}{0}{-1},
\frac{1}{\sqrt{2}}\mtrx{i}{0}{0}{i},
\frac{1}{\sqrt{2}}\mtrx{0}{1}{1}{0},
\frac{1}{\sqrt{2}}\mtrx{0}{-i}{i}{0}.
\]
Denote these four base matrices as $\betaM _k$, $k=0,1, 2, 3$, and
the first $16$ codes matrices can be represented by the linear
combinations $\sum _{k=0}^{3}{\chi_k \betaM _k}$. Similarly the
other $16$ code matrices, ${\mathcal C}_i$ ($16\le i \le 31$), are
of the form $\left[\begin{matrix}
A & -B^{\ast}\\
B & A^{\ast}
\end{matrix} \right]$, and can be represented with linear
combinations of four different base matrices $\betaM '_k$, $k=0,1,
2, 3$:
\[
\frac{1}{\sqrt{2}}\mtrx{1}{0}{0}{1},
\frac{1}{\sqrt{2}}\mtrx{i}{0}{0}{-i},
\frac{1}{\sqrt{2}}\mtrx{0}{-1}{1}{0},
\frac{1}{\sqrt{2}}\mtrx{0}{i}{i}{0}.
\]
It can be verified that $\betaM_k$'s satisfy Eq.~(\ref{nsc}) and
so do the $\betaM'_k$'s. However, it can be shown that the
property does not necessarily hold when two matrices are from two
different groups. The latter generator set is obtained from the
former via $\betaM'_k=\betaM_k \bd{U}$, $k=0, \ldots , 3$, where
$\bd{U} = \mtrx{1}{0}{0}{-1}$. Let ${\cal G}$ denote the first
$16$ codematrices, and ${\cal G}_{\rm e}$ all $32$ codematrices.
Clearly, ${\cal G}_{\rm e} = {\cal G} \cup {\cal G} \bd{U}$ and
${\cal G}_{\rm e}$ is a first-tier, direct, discernible expansion.
Thus, {\em all} $32$ matrices can be expressed as the linear
combinations of {\em eight} base matrices $\sum _{k=0}^{3}{\chi_k
\betaM _k}+\sum _{k=0}^{3}{\chi'_k \betaM '_k}$ where $\chi_k$ and
$\chi_k'$ ($k=0, 1, 2, 3$) are either $1$, $-1$, or $0$. Note that
either all $\chi_k$'s or all $\chi_k'$'s are zeros. That is,
\begin{multline}
\left\{{\mathcal C}_i\right\}_{i=0}^{31}=\left\{{\mathcal
C}_i\right\}_{i=0}^{15}\bigcup \left\{{\mathcal C}_i\right\}_{i=16}^{31}\\
=\left\{\sum _{k=0}^{3}{\left(\chi_k \betaM _k+\chi'_k
\betaM'_k\right)}: \chi_k \in \{-1, 1\} \mbox{ and } \chi_k'=0
\right\}
\\ \bigcup \left\{\sum _{k=0}^{3}{\left(\chi_k \betaM _k+\chi'_k
\betaM'_k\right)}: \chi'_k \in \{-1, 1\} \mbox{ and }
\chi_k=0\right\}.\nonumber \label{32matrices}
\end{multline}
We also remark that the space-time trellis codes in
\cite{IonescuMYL:01} are such that the branches departing from,
and converging to, any state are all labelled by codematrices from
either $\cal{G}$ or ${\cal G}\bd{U}$. As such, the side
information mentioned above is accessible to the decoder.
\bibliographystyle{ieeebib}
\bibliography{uniformity}

\begin{thebibliography}{99}
\bibitem{special} Special Issue on Space-Time Transmission, Reception, Coding and Signal Processing, \textit{IEEE Trans.\ Inform.\ Theory}, vol.\ 49, No.\ 10, October 2003.
\bibitem{ala:sim} S.\ M.\ Alamouti, ``A simple transmit diversity technique for wireless communications,'' \textit{IEEE     J.\ Select.\ Areas Commun.}, vol.\ 16, pp.\ 1451-1458, Oct.\ 1998.
\bibitem{tar:spa} V.\ Tarokh, N.\ Seshadri, and A.\ R.\ Calderbank, ``Space-time codes for high data rate wireless communication: Performance criteria and code construction,'' \textit{IEEE Trans.\ Inform.\ Theory}, vol.\ 44, No.\ 2, pp.\ 744-765, March 1998.
\bibitem{tar:ort} V.\ Tarokh, H.\ Jafarkhani, and A.\ R.\ Calderbank, ``Space-time block codes from orthogonal designs,'' \textit{IEEE Trans.\ Inform.\ Theory}, vol.\ 45, No.\ 5, pp.\ 1456-1467, July 1999.
\bibitem{sch:geo}H.\ Schulze, ``Geometrical Properties of Orthogonal Space-Time Codes,'' \textit{IEEE Commun.\ Letters}, vol.\ 7, pp.\ 64--66, Jan.\ 2003
\bibitem{gha:con} M.\ Gharavi-Alkhansari and A.\ B.\ Gershman, ''Constellation Space Invariance of Orthogonal Space-Time Block Codes,'' {\it IEEE Trans.\ Inform.\ Theory}, vol.\ 51, pp.\ 331--334, Jan.\ 2005.
\bibitem{yan:geo} Z.\ Yan and D.\ M.\ Ionescu, ''Geometrical Uniformity of  a Class of Space-Time Trellis Codes,'' {\it IEEE Trans.\ Inform.\ Theory}, vol.\ 50, pp.\ 3343--3347, Dec.\ 2004. 
\bibitem{tir:squ} O.\ Tirkkonen and A.\ Hottinen, ``Square-matrix embeddable space-time block codes for complex signal constellations,'' \textit{IEEE Trans.\ Inform.\ Theory}, vol.\ 48, pp.\ 384-395, Feb.\ 2002.
\bibitem{IonescuMYL:01} D.~M.\ Ionescu, K.~K.\ Mukkavilli, Z.~Yan, and J.~Lilleberg, ``{Improved 8- and 16-State Space-Time codes for 4PSK with Two Transmit Antennas},'' {\em {IEEE  Commun.\ Letters}}, vol.\ 5, pp.\ 301--303, July 2001.
\bibitem{ses:sup}N.\ Seshadri and H.\ Jafarkhani, ``Super-Orthogonal Space-Time Trellis Codes,'' \textit{Proc. ICC'02}, May 2002, Vol.\ 3, pp.\ 1439-1443.
\bibitem{jaf:sup} H.\ Jafarkhani, N.\ Seshadri, ``Super-orthogonal space-time trellis codes,'' \textit{IEEE Trans. Inform. Theory}, vol.\ 49, pp.\ 937-950, Apr.\ 2003.
\bibitem{siw:impcon} S.\ Siwamogsatham and M.\ P.\ Fitz, ``Improved High-Rate Space-Time Codes via Concatenation of Expanded Orthogonal Block Code and M-TCM,'' \textit{Proceedings of 2002 ICC}, vol. 1, pp.\ 636-640,
May 2002.
\bibitem{siw:import} S.\ Siwamogsatham and M.\ P.\ Fitz, ``Improved High-Rate Space-Time Codes via Orthogonality and Set Partitioning,'' \textit{Proceedings of 2002 IEEE WCNC}, vol. 1, pp.\ 264-270, March
2002.
\bibitem{big:per} E.\ Biglieri, G.\ Taricco, A.\ Tulino, ``Performance of space-time codes for a large number of antennas,'' \textit{IEEE Trans. Inform. Theory}, vol.\ 48, pp.\ 1794-1803, July 2002.
\bibitem{LWKC:03} H.-F.~Lu, Y.\ Wang, P.~V.~Kumar, and K.~M.~Chugg, ``{Remarks on Space-Time Codes Including a New Lower Bound and an Improved Code},'' \textit{IEEE Trans.\ Inform.\ Theory}, vol.\ 49, No.\ 10, pp.\ 2752-2757, October 2003.
\bibitem{Ionescu:03} D.\ M.\ Ionescu, ``On Space-Time Code Design,'' {\em IEEE Trans.\ Wireless Commun.}, vol.\ 2, pp.\ 20-28, Jan.\ 2003.
\end{thebibliography}

\end{document}